\documentclass{aa}
\usepackage{colortbl} \usepackage{multirow} \usepackage{dcolumn} \usepackage{amsmath}
\usepackage{amssymb} \usepackage{graphicx} \usepackage{epsfig} \usepackage{changebar}
\usepackage{natbib} \usepackage{txfonts} \usepackage{relsize}

\usepackage[figuresright]{rotating}

\newcolumntype{d}[1]{D{.}{\cdot}{#1}}

\newcolumntype{.}{D{.}{.}{-1}}

  \newcommand{\lsun}{L$_\odot$}

 \newcommand{\mum}{$\mu$m} 
   \newcommand{\kms}{km~s$^{-1}$}

\newcommand{\photons}{photon~cm$^{-2}$~s$^{-1}$} 

\begin{document} %
\title{Observational study of sites of triggered star formation}

\subtitle{CO and mid-infrared observations\thanks{Fig.~9 is only available in electronic form at \texttt{http:/www.aanda.org}}}

\author{J.~S.~Urquhart \inst{1,2} \and L.~K.~Morgan \inst{3,4} \and M.~A.~Thompson \inst{5}}

\offprints{J. S. Urquhart: james.urquhart@csiro.au}

\institute{School of Physics and Astronomy, University of Leeds, Leeds, LS2~9JT, UK \and Australia Telescope National Facility, CSIRO, Marsfield NSW 2122, Australia \and Green Bank
Telescope, P.O. Box 2, Green Bank, WV 24944, USA 
\and Department of Astronomy and Physics, Saint Mary's University, Halifax, NS B3H 3C3, Canada
\and Centre for Astrophysics Research, Science and
Technology Research Institute, University of Hertfordshire, College Lane, Hatfield, AL10 9AB, UK }

\date{}

\abstract 
{Bright-rimmed clouds (BRCs) are isolated molecular clouds located on the edges of evolved HII
regions. Star formation within these clouds may have been triggered through the propagation of
photoionisation-induced shocks driven by the expansion of the HII region.} 
{The main focus of this paper is to investigate the current level of star formation within a sample 
of southern hemisphere BRCs and evaluate to what extent, if any, star formation may have been triggered.} 
{In this paper we present the results of a programme of position-switched CO observations towards 45 southern BRCs. The $^{12}$CO, $^{13}$CO and C$^{18}$O $J$=1--0 rotational transitions were simultaneously observed using the 22-m Mopra telescope.
We complement these observations with archival mid-infrared data obtained from the MSX and Spitzer, as well as submillimetre and radio data previously reported in the literature. 
Combining all of the available data with the observations presented here allows us to build up a
comprehensive picture of the current level of star formation activity within a significant number of
BRCs.} 
{Analysis of the CO, mid-infrared and radio data result in the clouds being divided into three distinct groups: a) clouds that appear to be relatively unaffected by the photoionisation from the nearby OB star(s); b) clouds that show evidence of significant interaction between the molecular material and the HII regions; c) clouds towards which no CO emission is detected, but are associated with strong ionisation fronts; these are thought to be examples of clouds undergoing an ionisation flash. We refer to these groups as spontaneous, triggered, and zapped clouds, respectively.  Comparing the physical parameters of spontaneous and triggered samples we find striking differences in luminosity, surface temperature and column density with all three quantities significantly enhanced for the clouds considered to have been triggered. Furthermore, we find strong evidence for star formation within the triggered sample by way of  methanol and H$_2$O masers, embedded mid-infrared point sources and CO wings, however, we find evidence of ongoing star formation within only two of the spontaneous sample.  
} 
{We have used CO, mid-infrared and radio data to identify 24 of the 45 southern BRCs that are undergoing a strong interaction with their HII region. We can therefore exclude the other 21 sources ($\sim$50\%) from future studies of triggered star formation. Fourteen of the 24 interacting BRCs are found to be associated with embedded mid-infrared point sources and we find strong evidence of that these clouds are forming stars. The absence of mid-infrared sources towards the remaining ten clouds and the lack of any other evidence of star formation within these clouds leads us to conclude that these represent an earlier evolutionary stage of star formation. } 


\authorrunning{J. S. Urquhart et al.} 
\titlerunning{CO observations of BRCs} 
\maketitle
\section{Introduction} 
\subsection{Background}

Bright-rimmed clouds (BRCs) are isolated molecular clouds located on the edges of evolved HII regions. They are thought to result from the fragmentation of the dense layer of material swept up by the expanding HII region (known as  collect and collapse), or reflect density enhancements in the surrounding
clumpy medium. As the forming nebula surrounding a star ionising a HII region continues to expand it will preferentially erode away the low-density material, uncovering the higher density clumps, which begins to protrude into the HII
region. Once a clump is exposed to the HII region, the ionising radiation can have a dramatic affect on a cloud's structure and future evolution. Far-UV photons emitted from the nearby OB stars ionise the surface layers and drive shocks into the clouds. The process of star formation through the compression of a molecular cloud via a photoionisation-induced shock is known as radiatively-driven implosion (RDI; \citealt{bertoldi1989,lefloch1994,miao2006}). 

Their relative isolation and simple geometry make BRCs an ideal laboratory to investigate the RDI mode of triggered star formation, and estimate the contribution of RDI to the overall observed Galactic initial mass function. In order to investigate this mode of star formation Sugitani and collaborators (\citealt{sugitani1991, sugitani1994}) compiled  a catalogue of 89 BRCs (commonly referred to as the SFO catalogue) spread throughout the Galaxy; 44 clouds located in the northern hemisphere and 45 clouds located in the southern hemisphere. These clouds were identified by correlating IRAS point sources -- having colours consistent with embedded protostars -- with clouds displaying optically bright rims from the Sharpless HII region catalogue (\citealt{sharpless1959}) and the ESO(R) Southern Hemisphere Atlas.

The IRAS point sources associated with BRCs are systematically much more luminous than IRAS sources associated with more isolated clouds and Bok globules, in fact approximately three orders of magnitude more luminous (Sugitani et al. 1991, 1994). This has led to the suggestion that the RDI mode of star formation may preferentially lead to the formation of more massive stars, or small clusters of intermediate mass stars (e.g., \citealt{dobashi2001,morgan2008}). Moreover, Sugitani et al. found systematically higher luminosity to mass ratios for IRAS sources within BRCs than for isolated clouds, suggesting that RDI may also lead to a higher star formation efficiency than would otherwise be observed.

\subsection{Individual sources and systematic studies} 

To date, there have been a number of observational studies of small sub-samples of the SFO catalogue (e.g., \citealt{devries2002,urquhart2006a}) and some more detailed studies of individual sources (e.g., \citealt{lefloch1997,thompson2004a,urquhart2004,urquhart2007d}). These studies, in general, reveal the presence of recent or ongoing star formation within the small number of BRCs studied, as well as strong circumstantial evidence which supports the hypothesis that the observed star formation has been triggered. Although we have learnt a great deal about star formation from these studies, they have told us little about whether star formation is common within BRCs, what portion is likely to have been triggered, or allowed us to evaluate the efficiency of this mode of star formation.

In order to address some of these more general questions a small number of large observational studies have been conducted. These have been aimed at investigating the properties of the entire sample of BRCs in the SFO catalogue. Radio continuum (cm) data has been presented for the whole of the catalogue (all 89 clouds) in two papers (\citealt{thompson2004b,morgan2004}). These focused on mapping the distribution of ionised gas associated with the optically bright rims, and determining the pressure balance between the ionised-molecular gas interface. These studies detected radio emission towards 51 clouds ($\sim$60\%), clearly indicating that a significant number of clouds are being photoionised.

More recently, \citealt{morgan2008} reported the results of a set of submillimetre observations made towards 45 BRCs using the SCUBA on the James Clerk Maxwell Telescope (JCMT). The sources observed are primarily located within the northern hemisphere, however, they included six southern hemisphere sources observable from the JCMT. These
observations revealed the presence of dense cores located within almost all observed BRCs, with temperatures, and luminosities, consistent with the presence of ongoing star formation. Many of these clouds are also associated with radio emission, clearly demonstrating an interaction between the HII region and the BRCs, and thus, consistent with the hypothesis that the star formation could have been induced by RDI.

Detailed investigations of a handful of BRCs indicate that they are forming stars via the RDI mechanism, or contain dense cores which have properties consistent with those of protostellar cores. However, the causative link between the external influence of ionising radiation on these clouds and the star formation within has not yet been meaningfully explored across the whole sample.
Moreover, the non-detection of any radio emission towards $\sim$40\% of the sample and submm emission towards 10\% of sources observed with SCUBA would suggest that the sample includes a significant number of sources which do not represent a mode of triggered star formation. These contaminating sources need to be removed before any meaningful statistical results can be obtained. In this paper we will combine molecular line observations, cm radio continuum and mid-infrared imaging archival data to investigate the level of star formation across the whole southern sample. Our primary aim is to identify the clouds in which star formation is likely to have been triggered, and eliminate clouds in which star formation is unlikely to have been triggered from future studies.

\subsection{Radiatively-Driven Implosion: structure and evolution of BRCs} \label{sect:rdi}

There has been a considerable amount of theoretical work presented in the literature that deals specifically with the impact of ionising radiation on a BRC exposed to a HII region (e.g., \citealt{bertoldi1989,bertoldi1990,lefloch1994,kessel2003,miao2006}). The pressure balance between the external hot ionised gas, and the cooler molecular gas within the cloud has emerged as a key diagnostic that can be used to evaluate the impact the arrival of an ionisation front has on the dynamics and future evolution of a cloud.  We summarise the findings of the theoretical models here to aid later discussion of the classification of the clouds.

The numerical models predict three distinct outcomes of a cloud's exposure to the strong far-UV radiation field of the HII region: 1) the ionisation photon flux is too low and/or the pressure of the ionised gas is too small to compress the molecular gas, the radiation field is therefore unable to dynamically affect the cloud; 2) the far-UV photon flux is too strong and/or the density of the molecular material is too small, the ionisation front is able to propagate supersonically with respect to the molecular gas, resulting in an almost instantaneous photoionisation of the cloud; this is referred to as an \emph{ionisation flash} by \citet{lefloch1994} or \emph{cloud zapping} by \citet{bertoldi1989}; 3) the ionisation flux and the density of the molecular cloud is such that the whole evolution of the cloud is dominated by the propagation of a D-critical ionisation front that also acts to drive an isothermal shock before it.

It is this last possible outcome that is of most interest to us in terms of determining which clouds are likely to host triggered star
formation. According to the models, this outcome is the most common for a wide range of possible input parameters. From our theoretical
understanding and recent detailed multi-wavelength observational studies of BRCs and cometary globules (e.g.,
\citealt{lefloch2002,deharveng2003,rho2006,urquhart2007d,lefloch2008})  we have formed the following picture of their structure and evolution.

\noindent Once a cloud is exposed to ionising radiation an ionisation front forms over its exposed surface. Assuming that the pressure of the ionised gas exceeds that of the nebula, and the molecular gas within the BRC, it will begin to expand into the HII region leading to the formation of a photoionised layer of gas that surrounds the cloud, often referred to as an \emph{ionised boundary layer} (IBL). Recombination within the IBL acts to limit the amount of fresh material ionised by the far-UV photon flux. However, a significant fraction of the UV radiation manages to propagate through the IBL into the surface layers and begins to dissociate the molecular gas, leading to the formation of a photon-dominated region (PDR). The over-pressure of the hot, ionised surface layers with respect to the cooler neutral gas leads to isothermal shocks being driven into the cloud. These shocks compress the molecular gas and can lead to the formation, and trigger the subsequent collapse, of dense cores, as well as triggering the collapse of subcritical cores that pre-date the arrival of the ionisation front (\citealt{elmegreen1992}). 

The three main regions we will be focusing on in this paper are (moving from the HII region into the cloud) the IBL, the PDR and the internal molecular gas within the cloud. These three regions can be studied using a combination of radio continuum observations to trace the ionised gas, mid-infrared imaging to map the PDR, and CO line emission to probe the internal structure of the cloud. Combining these three complementary data sets allow us evaluate the impact the ionisation fronts are having on the evolution of the 45 BRCs in the southern catalogue of \cite{sugitani1994}.

The structure of this paper is as follows: in Sect.~\ref{sect:observations} we briefly describe the CO observations, data reduction and derived physical quantities for each cloud. Complementary archival data are introduced in Sect.~\ref{sect:archival}. Our results and analyses are presented in Sect.~\ref{sect:results} followed by a discussion in Sect.~\ref{sect:discussion}. We present a summary and highlight our main findings in Sect.~\ref{sect:summary}.

\section{Observations}
\label{sect:observations}

\subsection{Southern catalogue of BRCs} \label{sect:brc_summary}

\begin{figure*} \begin{center}

\includegraphics[height=0.95\linewidth, angle=270, trim=0 20 0 50]{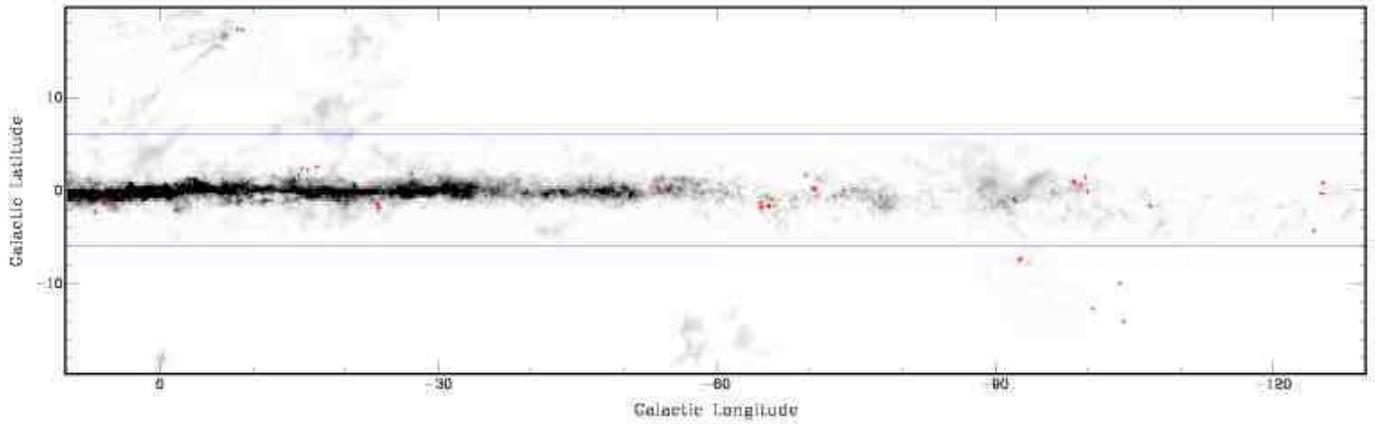}

\caption{\label{fig:brc_dist} Galactic distribution of the southern catalogue of BRCs. The greyscale shows the distribution of molecular gas as traced by the integrated $^{12}$CO map of \citet{dame2001}. The positions of the southern sample of BRCs are indicated by red crosses. The area between the blue lines indicates the region covered by the MSX survey (see text for details).} \end{center}

\end{figure*}

\renewcommand{\thefootnote}{\alph{footnote}}
\setcounter{footnote}{0}
\begin{table*} 
\begin{minipage}{\textwidth} 

\begin{center} \small \caption{\label{tbl:hii_regions}List of HII regions and ionising stars associated with southern hemisphere bright-rimmed clouds (listed in order of increasing Galactic longitude). } \medskip
\label{tbl:ionising_stars} \begin{tabular}{lllll}\hline\hline HII region\footnotemark \setcounter{footnote}{0} & Ionising star(s) &
Spectral Type & $D$ (kpc) & Associated BRCs \\\hline 
S 307 		& 	MFJ 3 			& 	B0 V 		& 	3.6 	& 	SFO 49\\ 
S 306 		&	LSS 458 		& 	O5 			& 	4.2 	& 	SFO 47\\ 
			& 	LSS 467 		& 	O9 III 		& 	& \\ 
RCW 14 		& 	HD 57236 		& 	O8 V 		& 	1.93 	& 	SFO 45\\ 
Gum Nebula 	& $\zeta$ Pup 		& 	O4 If 		& 	0.45 	& 	SFO 46, SFO 48,\\
			& $\gamma^{2}$ Vel 	& 	WC 8+O8 III & 			&	SFO 50--53\\ 
NGC 2626 	& 	vBH 17a 		& 	B1 V 		& 	0.95 	& 	SFO 54\\ 
RCW 27 		& 	HD 73882 		& 	O8 V 		& 	1.15	& 	SFO 55,SFO 56\\ 
RCW 32 		& 	HD 74804 		& 	B0 V--B4 II & 	0.70 	& 	SFO 57, SFO 58\\ 
RCW 38 		& 	RCW 38/IRS 2 	& 	O5 			& 	1.7		& 	SFO 59, SFO 60\\ 
NGC 3503 	& 	vBH 46a 		& 	B0 Ve 		& 	2.88 	& 	SFO 61--63\\ %
			& 	HD 305938 		& 	B1 V 		& 			& \\
BBW 347 	& 	LSS 2231 		& 	B0 V 		& 	2.7 	& 	SFO 64\\ 
RCW 62 		& 	HD 101131 		& 	O6.5 n 		& 	1.7 	& 	SFO 65--70\\ 
			& 	HD 101205		& 	O6.5 		& 			& \\ 
			& 	HD 101436 		& 	O7.5 		& 			& \\ 
(Cen R 1) 	& 	vBH 59 			& 	B6 I 		& 	2.0 	& 	SFO 71\\ 
RCW 75 		& 	HD 115455 		& 	O7.5 III 	& 	1.9 	& 	SFO 72, SFO 73\\ RCW 85 		& 	HD 124314 		& 	O6 V 		& 	1.5 	& SFO 74\\ 
RCW 98 		& 	LSS 3423 		& 	O9.5 IV 	& 	2.8 	& SFO 75\\ 
RCW 105 	& 	HD 144918 		& 	O7 			& 	1.8 	& SFO 76\\ 
($\sigma$ Sco) & HD 147165		& 	B1 III 		& 	0.15 	& SFO 77, SFO 78 \\
			& CCDM~J16212$-$2536AB & B1 		& 			& \\ 
RCW 108 	& 	HD 150136 		& 	O5 III		& 	1.35 	& SFO 79--81\\ 
			& 	HD 150135 		& 	O6.5 V 		& 			& \\ 
RCW 113/116 & 	HD 152233 		&	O6 III 		& 	1.85 	& SFO 82--85\\ 
			& 	HD 326286 		&	B0	 		&		 	& \\ 
			& 	HD 152245 		&	B0 Ib 		& 		 	& \\ 
RCW 134 	& 	HD 161853 		& 	O8 V 		& 	1.6 	& SFO 86\\ 
M 8 		& 	HD 164794 		& 	O4 V 		& 	1.86 	& SFO 87, SFO 88\\ 
S 29 		& 	HD 165921 		& 	O7.5 V 		& 	1.86 	& SFO 89\\ 
\hline \end{tabular} 

\footnotetext[1]{Parentheses indicate that the listed object is a star rather than an HII region.}

\end{center}
\end{minipage}
\end{table*}

\renewcommand{\thefootnote}{\arabic{footnote}}
\setcounter{footnote}{0}

The 45 BRCs in the southern catalogue are associated with 22 HII regions. These are located at heliocentric distances ranging from 0.15--4.20~kpc, with a mean distance of $\sim$2~kpc. The sample has a  Galactic longitude distribution ($l$) between $-$120 and 10\degr\ with a latitude distribution $|b|$ $<$ 18\degr, however, the majority of the clouds are located within the Galactic disk (37 of the 45 BRCs have $|b| <$ 3\degr). The HII regions, their ionising star(s), heliocentric distances and associated BRCs are listed in Table~1. The ionising star(s), spectral types and assumed distances are drawn from \citet{yamaguchi1999} and  \citet{thompson2004a} and references therein, apart from $\sigma$ Sco where the spectral type and assumed distance are taken from \citet{eggen1998}. The spectral types of HD~305938, CCDM~J16212$-$2536AB, HD~326286 and HD~152245 have been drawn from SIMBAD. The SFO identification number, corresponding IRAS name and position in Galactic coordinates are given for each cloud in Table~2 and in Fig.~\ref{fig:brc_dist} we present a plot of their distribution with respect to the Galactic plane.

\begin{table} \begin{center} \caption{\citet{sugitani1994} southern BRC catalogue.} \small
\label{tbl:brc_catalogue} \begin{tabular}{lc..}\hline\hline Cat Id. &   IRAS Name   &   l        
& b \\ \hline SFO 45	&	07162-2200	&	235.6090	&	-4.2905	\\ SFO
46$^\star$	&	07178-4429	&	256.1510	&	-14.0548	\\ SFO
47	&	07296-1921	&	234.7610	&	-0.2766	\\ SFO
48$^\star$	&	07329-4647	&	259.4560	&	-12.6856	\\ SFO
49	&	07334-1842	&	234.6410	&	0.8337	\\ SFO
50$^\star$	&	07388-4259	&	256.5090	&	-9.9985	\\ SFO
51	&	08076-3556	&	253.2930	&	-1.6127	\\ SFO
52$^\star$	&	08242-5050	&	267.3660	&	-7.4970	\\ SFO
53$^\star$	&	08250-5030	&	267.1600	&	-7.2026	\\ SFO
54	&	08337-4028	&	259.9410	&	-0.0405	\\ SFO
55	&	08393-4041	&	260.7750	&	0.6783	\\ SFO
56	&	08411-3949	&	260.2980	&	1.4790	\\ SFO
57	&	08423-4105	&	261.4310	&	0.8619	\\ SFO
58	&	08435-4105	&	261.5790	&	1.0546	\\ SFO
59	&	08563-4711	&	267.7280	&	-1.1022	\\ SFO
60	&	08583-4719	&	268.0560	&	-0.9499	\\ SFO
61	&	10581-5920	&	289.2800	&	0.2888	\\ SFO
62	&	10591-5934	&	289.5020	&	0.1172	\\ SFO
63	&	11012-5931	&	289.7170	&	0.2680	\\ SFO
64	&	11101-5829	&	290.3740	&	1.6612	\\ SFO
65	&	11306-6311	&	295.2960	&	-1.9188	\\ SFO
66	&	11315-6259	&	294.3350	&	-1.7033	\\ SFO
67	&	11317-6254	&	294.3310	&	-1.6172	\\ SFO
68	&	11332-6258	&	294.5120	&	-1.6235	\\ SFO
69	&	11388-6306	&	295.1600	&	-1.5802	\\ SFO
70	&	11398-6251	&	295.2000	&	-1.3031	\\ SFO
71	&	13050-6154	&	304.8880	&	0.6341	\\ SFO
72	&	13158-6217	&	306.1160	&	0.1382	\\ SFO
73	&	13168-6208	&	306.2440	&	0.2865	\\ SFO
74	&	14159-6111	&	313.2850	&	-0.3365	\\ SFO
75	&	15519-5430	&	327.5740	&	-0.8517	\\ SFO
76	&	16069-4858	&	332.9560	&	1.8038	\\ SFO
77$^\star$	&	16168-2526	&	351.1320	&	17.2349	\\ SFO
78$^\star$	&	16178-2501	&	351.6170	&	17.3588	\\ SFO
79	&	16362-4845	&	336.4910	&	-1.4751	\\ SFO
80	&	16365-4836	&	336.6380	&	-1.4060	\\ SFO
81	&	16373-4911	&	336.2910	&	-1.9026	\\ SFO
82	&	16438-4110	&	343.0840	&	2.5238	\\ SFO
83	&	16487-4043	&	344.0230	&	2.1170	\\ SFO
84	&	16502-4002	&	344.7360	&	2.3186	\\ SFO
85	&	16555-4237	&	343.3520	&	-0.0769	\\ SFO
86	&	17463-3128	&	358.2610	&	-2.0554	\\ SFO
87	&	17597-2422	&	5.8882	&	-1.0079	\\ SFO
88	&	18012-2407	&	6.2673	&	-1.1657	\\ SFO
89	&	18068-2405	&	6.9217	&	-2.2707	\\ \hline \end{tabular} \end{center}
\noindent $^\star$ Indicates that the BRC is located outside the MSX survey region. \end{table}

\subsection{Mopra millimetre line observations}

Millimetre CO ($J$=1--0) observations were made using the Mopra 22m telescope which is located near Coonabarabran, New South Wales, Australia.\footnote{Mopra is operated by the Australia Telescope National Facility, CSIRO and the University of New South Wales.} The telescope is situated at an elevation of 866 metres above sea level, and at a latitude of 31 degrees south.

The receiver is a cryogenically cooled \mbox{($\sim$ 4 K)}, low-noise,
superconductor-insulator-superconductor (SIS) junction mixer with a frequency range between 85--116 GHz, corresponding to a half-power beam-width of \mbox{36\arcsec--33\arcsec} (Mopra Technical Summary\footnote{Available at \texttt{http://www.narrabri.atnf.csiro.au/mopra/.}}). The spectrometer is made
up of four 2.2~GHz bands which overlap slightly to provide a total of 8~GHz continuous bandwidth. Up to four zoom windows can be placed within each 2.2~GHz band allowing up to 16 spectral lines to be observed simultaneously. Each zoom window provides a bandwidth of 267 MHz with 4096 channels, corresponding to a total velocity range of $\sim$450~\kms\ and resolution of $\sim$0.1 \kms\ per
channel.

We conducted position-switched observations towards all 45 southern hemisphere BRCs, centred on the position of the IRAS point source associated with each cloud (see Table~\ref{tbl:brc_catalogue} for positions). We used three of the zoom windows to observe the $^{12}$CO, $^{13}$CO and C$^{18}$O ($J$=1--0) transitions simultaneously. Each source was observed for a total of five minutes of on-source integration, split into a number of separate scans consisting of 1 minute on- and 1 minute off-source. Typical rms values of $\sim$0.1 K per channel were obtained for the $^{13}$CO and C$^{18}$O transitions and $\sim$0.3~K per channel for the $^{12}$CO transition. Reference positions were offset from the BRC positions by 1 degree in a direction perpendicular to the
Galactic plane. These were chosen to avoid contamination of source emission from emission in the reference position at a similar velocity. In some cases, particularly towards the Galactic centre, several positions needed to be tried before a suitable reference position was found.

Telescope pointing was checked approximately every hour by observing a nearby SiO maser and was found to be better than 10\arcsec. To correct the measured antenna temperatures ($T_{\rm{A}}$) for atmospheric absorption, ohmic losses and rearward spillover, a measurement was made of an ambient load (assumed to be at 290~K) following the method of \citet{kutner1981}; this was done before each source was observed. System temperatures were found to be stable during our observations; $T_{\rm{sys}}$ measurements were between 250--300~K for the $^{13}$CO and C$^{18}$O transitions and 500--600 K for the $^{12}$CO transition. Absolute calibration was performed by comparing measured line temperatures of Orion KL and M17SW to standard values. We estimate the combined calibration uncertainties to be no more than 20\%.

The data were reduced using the ATNF spectral line reducing software (ASAP). The reduction steps consisted of quotienting the individual on-off scans to remove sky emission (these scans were then inspected and poor scans were removed), fitting a low-order polynomial to the baseline, before averaging the scans together to produce a single spectrum for each transition, for each source. Finally the spectra were calibrated to the corrected main beam brightness temperature scale ($T_{\rm{MB}}^*$) by dividing the corrected antenna temperatures by the telescope efficiency where $T_{\rm{MB}}^*=T^*_{\rm{A}}/\eta_{\rm{MB}}$.  The main beam efficiency of the telescope at 115~GHz is $\sim$0.42, however, since the emission associated with our sources is expected to be extended (BRCs have typical sizes of a few arcmins) they are likely to couple well to the main and inner error beams. We have therefore used the extended main beam efficiency of $\sim$0.55 to take account of the contribution of the inner error beam (see \citealt{ladd2005} for details).

Spectral line parameters were obtained by fitting Gaussian profiles to each component present in a given spectrum using the Onsala Space Observatory spectral line analysis package \textsl{XS}.\footnote{For more details see \texttt{\hyphenchar\font45\sloppy http://www.chalmers.se/rss/oso-en/observations/data-reduction-software}.} Where necessary, higher-order polynomials were fitted and subtracted from the baselines beforehand. We present the parameters obtained from the Gaussian fits to these data and derived values in Table~\ref{tbl:co_results} and present plots of the fitted spectra in Fig.~\ref{fig:fitted_spectra}.

\begin{table*} 
\tiny 
\begin{center} 
\caption{CO ($J$=1--0) line parameters obtained from Gaussian fits to the spectra. Non-detections in a relevant transition are indicated by $\cdots$.} 
\label{tbl:co_results}

\begin{tabular}{@{}l...c...c...@{}} \hline \hline &\multicolumn{3}{c}{$^{12}$CO} &&
\multicolumn{3}{c}{$^{13}$CO} && \multicolumn{3}{c}{C$^{18}$O} \\ \cline{2-4} \cline{6-8}
\cline{10-12} & \multicolumn{1}{c}{$v$}& \multicolumn{1}{c}{$T_R^*$} & \multicolumn{1}{c}{$\Delta
v$} && \multicolumn{1}{c}{$v$}&\multicolumn{1}{c}{$T_R^*$} & \multicolumn{1}{c}{$\Delta v$} &&
\multicolumn{1}{c}{$v$}&\multicolumn{1}{c}{$T_R^*$} & \multicolumn{1}{c}{$\Delta v$} \\ SFO Id. &
\multicolumn{1}{c}{(km s$^{-1}$)}& \multicolumn{1}{c}{(K)} & \multicolumn{1}{c}{(km s$^{-1}$)}  &&
\multicolumn{1}{c}{(km s$^{-1}$)}&\multicolumn{1}{c}{(K)} & \multicolumn{1}{c}{(km s$^{-1}$)}  &&
\multicolumn{1}{c}{(km s$^{-1}$)}&\multicolumn{1}{c}{(K)} & \multicolumn{1}{c}{(km s$^{-1}$)}  \\
\hline 45 &23.4&10.1&1.9& &23.5&2.4&1.3&
&\multicolumn{1}{c}{$\cdots$}&\multicolumn{1}{c}{$\cdots$}&\multicolumn{1}{c}{$\cdots$} \\
46&3.9&10.7&1.2& &3.6&2.1&0.5&
&\multicolumn{1}{c}{$\cdots$}&\multicolumn{1}{c}{$\cdots$}&\multicolumn{1}{c}{$\cdots$} \\ 
47&43.4&21.3&1.3& &43.5&2.5&1.0&
&\multicolumn{1}{c}{$\cdots$}&\multicolumn{1}{c}{$\cdots$}&\multicolumn{1}{c}{$\cdots$} \\
48&1.4&4.8&1.0& &1.4&1.2&0.9&
&\multicolumn{1}{c}{$\cdots$}&\multicolumn{1}{c}{$\cdots$}&\multicolumn{1}{c}{$\cdots$} \\ 
49&47.1&22.5&2.6& &47.0&4.8&1.4&
&\multicolumn{1}{c}{$\cdots$}&\multicolumn{1}{c}{$\cdots$}&\multicolumn{1}{c}{$\cdots$} \\
50&-4.3&2.9&1.6&
&\multicolumn{1}{c}{$\cdots$}&\multicolumn{1}{c}{$\cdots$}&\multicolumn{1}{c}{$\cdots$}&
&\multicolumn{1}{c}{$\cdots$}&\multicolumn{1}{c}{$\cdots$}&\multicolumn{1}{c}{$\cdots$} \\ 
51&5.8&9.9&2.1& &6.1&6.4&1.1& &6.1&1.5&0.7 \\ &7.5&2.6&7.7&
&\multicolumn{1}{c}{$\cdots$}&\multicolumn{1}{c}{$\cdots$}&\multicolumn{1}{c}{$\cdots$}&
&\multicolumn{1}{c}{$\cdots$}&\multicolumn{1}{c}{$\cdots$}&\multicolumn{1}{c}{$\cdots$} \\
52&5.3&14.5&1.3& &5.3&8.0&0.8& &5.3&1.9&0.7 \\ 
53&4.8&14.0&1.2& &4.8&7.2&0.9& &4.8&1.5&0.6 \\ 
54&8.0&19.0&6.3& &7.8&8.0&2.4& &7.8&1.9&1.9 \\ 
55 &7.6&31.5&2.6& &7.4&8.2&1.7& &7.3&1.7&1.2\\
56&5.2&10.4&1.3&
&\multicolumn{1}{c}{$\cdots$}&\multicolumn{1}{c}{$\cdots$}&\multicolumn{1}{c}{$\cdots$}&
&\multicolumn{1}{c}{$\cdots$}&\multicolumn{1}{c}{$\cdots$}&\multicolumn{1}{c}{$\cdots$} \\ 
57&5.3&28.5&2.7& &5.1&7.6&2.2& &4.6&1.0&1.9 \\ 
58 &4.4&31.0&2.0& &4.4&8.5&1.4& &4.3&1.2&1.3  \\
59&5.7&11.4&2.1& &5.3&2.5&1.6&
&\multicolumn{1}{c}{$\cdots$}&\multicolumn{1}{c}{$\cdots$}&\multicolumn{1}{c}{$\cdots$} \\
&1.5&4.6&2.2&
&\multicolumn{1}{c}{$\cdots$}&\multicolumn{1}{c}{$\cdots$}&\multicolumn{1}{c}{$\cdots$}&
&\multicolumn{1}{c}{$\cdots$}&\multicolumn{1}{c}{$\cdots$}&\multicolumn{1}{c}{$\cdots$} \\
60&7.8&6.5&2.5& &6.5&0.7&3.7&
&\multicolumn{1}{c}{$\cdots$}&\multicolumn{1}{c}{$\cdots$}&\multicolumn{1}{c}{$\cdots$} \\
&-0.2&4.7&2.9&
&\multicolumn{1}{c}{$\cdots$}&\multicolumn{1}{c}{$\cdots$}&\multicolumn{1}{c}{$\cdots$}&
&\multicolumn{1}{c}{$\cdots$}&\multicolumn{1}{c}{$\cdots$}&\multicolumn{1}{c}{$\cdots$} \\
61&-23.1&3.3&2.9&
&\multicolumn{1}{c}{$\cdots$}&\multicolumn{1}{c}{$\cdots$}&\multicolumn{1}{c}{$\cdots$}&
&\multicolumn{1}{c}{$\cdots$}&\multicolumn{1}{c}{$\cdots$}&\multicolumn{1}{c}{$\cdots$} \\ 
62&20.7&5.5&4.1& &20.1&0.4&4.7&
&\multicolumn{1}{c}{$\cdots$}&\multicolumn{1}{c}{$\cdots$}&\multicolumn{1}{c}{$\cdots$} \\
&-25.6&20.4&2.3& &-25.7&2.8&2.1&
&\multicolumn{1}{c}{$\cdots$}&\multicolumn{1}{c}{$\cdots$}&\multicolumn{1}{c}{$\cdots$} \\
64&-18.4&20.7&4.4& &-18.5&4.5&2.7& &-18.4&0.7&2.3 \\ 
65 &-18.5&16.8&3.8&
&-18.7&3.8&2.9&
&\multicolumn{1}{c}{$\cdots$}&\multicolumn{1}{c}{$\cdots$}&\multicolumn{1}{c}{$\cdots$} \\ 
66 &-14.8&16.3&2.1& &-14.7&5.3&1.5& &-14.7&1.1&1.3 \\ 
67 &-14.8&26.1&2.8& &-14.5&7.8&1.9&&-14.4&1.6&1.3 \\ 
68 &-16.1&21.9&4.7& &-15.6&8.2&2.5& &-15.6&1.7&2.3 \\ 
69 &-12.5&16.5&2.1&&-12.6&1.4&2.0&
&\multicolumn{1}{c}{$\cdots$}&\multicolumn{1}{c}{$\cdots$}&\multicolumn{1}{c}{$\cdots$} \\
70&-15.6&14.7&2.1& &-16.1&1.6&1.7&
&\multicolumn{1}{c}{$\cdots$}&\multicolumn{1}{c}{$\cdots$}&\multicolumn{1}{c}{$\cdots$} \\
71&-35.6&16.7&3.3& &-35.8&7.1&2.5& &-36.1&1.5&1.9 \\ 
&-38.5&5.1&8.4&&\multicolumn{1}{c}{$\cdots$}&\multicolumn{1}{c}{$\cdots$}&\multicolumn{1}{c}{$\cdots$}&
&\multicolumn{1}{c}{$\cdots$}&\multicolumn{1}{c}{$\cdots$}&\multicolumn{1}{c}{$\cdots$}  \\
72&-32.9&27.7&2.1& &-32.8&7.1&1.6& &-32.7&1.0&1.6 \\ 
73&-28.6&13.2&2.7& &-28.7&3.5&1.7&&-28.8&0.5&1.3 \\ 
&-34.0&3.1&3.2&
&\multicolumn{1}{c}{$\cdots$}&\multicolumn{1}{c}{$\cdots$}&\multicolumn{1}{c}{$\cdots$}&
&\multicolumn{1}{c}{$\cdots$}&\multicolumn{1}{c}{$\cdots$}&\multicolumn{1}{c}{$\cdots$} \\
&-40.9&9.7&4.8& &-40.9&2.9&2.3& &-40.9&0.6&1.3 \\ 
&-47.7&3.3&2.2&
&\multicolumn{1}{c}{$\cdots$}&\multicolumn{1}{c}{$\cdots$}&\multicolumn{1}{c}{$\cdots$}&
&\multicolumn{1}{c}{$\cdots$}&\multicolumn{1}{c}{$\cdots$}&\multicolumn{1}{c}{$\cdots$} \\
74&-28.3&22.9&2.6& &-28.0&4.5&2.0& &-27.8&0.5&1.6 \\ 
75&-36.5&38.3&3.6& &-36.5&14.0&2.5& &-36.6&2.9&2.3 \\ 
76&-22.2&31.8&2.8& &-22.2&6.4&2.2& &-22.3&8.0&2.0 \\ 
79&-24.9&45.5&3.1& &-24.3&12.8&2.6& &-24.0&2.6&3.3 \\ &-21.4&20.6&4.8& &-21.7&5.4&3.3&
&\multicolumn{1}{c}{$\cdots$}&\multicolumn{1}{c}{$\cdots$}&\multicolumn{1}{c}{$\cdots$} \\
80&-25.4&10.8&2.6& &-24.8&1.6&2.2&
&\multicolumn{1}{c}{$\cdots$}&\multicolumn{1}{c}{$\cdots$}&\multicolumn{1}{c}{$\cdots$} \\
&-31.2&2.6&2.7&
&\multicolumn{1}{c}{$\cdots$}&\multicolumn{1}{c}{$\cdots$}&\multicolumn{1}{c}{$\cdots$}&
&\multicolumn{1}{c}{$\cdots$}&\multicolumn{1}{c}{$\cdots$}&\multicolumn{1}{c}{$\cdots$} \\
81&-10.1&4.9&1.0&
&\multicolumn{1}{c}{$\cdots$}&\multicolumn{1}{c}{$\cdots$}&\multicolumn{1}{c}{$\cdots$}&
&\multicolumn{1}{c}{$\cdots$}&\multicolumn{1}{c}{$\cdots$}&\multicolumn{1}{c}{$\cdots$} \\
&-18.9&11.4&1.4& &-18.9&2.3&0.9&
&\multicolumn{1}{c}{$\cdots$}&\multicolumn{1}{c}{$\cdots$}&\multicolumn{1}{c}{$\cdots$} \\
&-29.0&5.5&1.8&
&\multicolumn{1}{c}{$\cdots$}&\multicolumn{1}{c}{$\cdots$}&\multicolumn{1}{c}{$\cdots$}&
&\multicolumn{1}{c}{$\cdots$}&\multicolumn{1}{c}{$\cdots$}&\multicolumn{1}{c}{$\cdots$} \\
&-42.5&2.6&4.8&
&\multicolumn{1}{c}{$\cdots$}&\multicolumn{1}{c}{$\cdots$}&\multicolumn{1}{c}{$\cdots$}&
&\multicolumn{1}{c}{$\cdots$}&\multicolumn{1}{c}{$\cdots$}&\multicolumn{1}{c}{$\cdots$} \\
82&-24.2&28.1&3.1& &-24.1&8.9&2.0& &-24.2&1.7&1.7 \\ 83&-7.0&1.8&1.4&
&\multicolumn{1}{c}{$\cdots$}&\multicolumn{1}{c}{$\cdots$}&\multicolumn{1}{c}{$\cdots$}&
&\multicolumn{1}{c}{$\cdots$}&\multicolumn{1}{c}{$\cdots$}&\multicolumn{1}{c}{$\cdots$} \\
84&-10.1&17.0&1.9& &-10.2&2.3&1.7&
&\multicolumn{1}{c}{$\cdots$}&\multicolumn{1}{c}{$\cdots$}&\multicolumn{1}{c}{$\cdots$} \\
85&9.0&3.1&2.3&
&\multicolumn{1}{c}{$\cdots$}&\multicolumn{1}{c}{$\cdots$}&\multicolumn{1}{c}{$\cdots$}&
&\multicolumn{1}{c}{$\cdots$}&\multicolumn{1}{c}{$\cdots$}&\multicolumn{1}{c}{$\cdots$} \\
&-1.6&2.1&2.4&
&\multicolumn{1}{c}{$\cdots$}&\multicolumn{1}{c}{$\cdots$}&\multicolumn{1}{c}{$\cdots$}&
&\multicolumn{1}{c}{$\cdots$}&\multicolumn{1}{c}{$\cdots$}&\multicolumn{1}{c}{$\cdots$} \\
&-19.1&32.4&3.7& &-19.4&5.0&2.2& &-19.7&0.6&1.9 \\
&-25.1&4.7&3.7& &-25.1&0.6&8.6&
&\multicolumn{1}{c}{$\cdots$}&\multicolumn{1}{c}{$\cdots$}&\multicolumn{1}{c}{$\cdots$} \\
&-39.8&2.2&2.7& &-39.7&0.6&2.1&
&\multicolumn{1}{c}{$\cdots$}&\multicolumn{1}{c}{$\cdots$}&\multicolumn{1}{c}{$\cdots$} \\
86&11.4&3.7&2.3&
&\multicolumn{1}{c}{$\cdots$}&\multicolumn{1}{c}{$\cdots$}&\multicolumn{1}{c}{$\cdots$}&
&\multicolumn{1}{c}{$\cdots$}&\multicolumn{1}{c}{$\cdots$}&\multicolumn{1}{c}{$\cdots$} \\
&2.4&21.3&3.1& &2.1&6.7&2.1& &1.9&1.4&1.6 \\ &4.2&5.1&2.5& &3.9&3.8&1.3& &3.8&1.4&1.2 \\
&-1.3&7.7&2.5&
&\multicolumn{1}{c}{$\cdots$}&\multicolumn{1}{c}{$\cdots$}&\multicolumn{1}{c}{$\cdots$}&
&\multicolumn{1}{c}{$\cdots$}&\multicolumn{1}{c}{$\cdots$}&\multicolumn{1}{c}{$\cdots$} \\
87&12.5&33.5&1.9& &12.5&9.3&1.1& &12.5&1.8&0.8 \\ &15.6&23.4&1.9& &15.6&3.8&1.5& &15.7&0.5&1.0 \\
88&13.9&26.5&4.2& &14.6&3.5&1.9&
&\multicolumn{1}{c}{$\cdots$}&\multicolumn{1}{c}{$\cdots$}&\multicolumn{1}{c}{$\cdots$} \\
&13.9&26.5&4.2& &13.1&1.9&2.7&
&\multicolumn{1}{c}{$\cdots$}&\multicolumn{1}{c}{$\cdots$}&\multicolumn{1}{c}{$\cdots$} \\
89&10.3&7.9&2.0& &10.6&0.8&1.9&
&\multicolumn{1}{c}{$\cdots$}&\multicolumn{1}{c}{$\cdots$}&\multicolumn{1}{c}{$\cdots$} \\

\hline
\end{tabular}
\end{center}

\end{table*}

\section{Archival data}\label{sect:archival} \subsection{Mid-infrared archival data: MSX \& Spitzer } \label{sect:msx}

The Midcourse Space Experiment (MSX) satellite surveyed the whole Galactic plane with $|b|$ $<$ 6\degr\ in four wavelength bands between 6 and 25 $\mu$m with a resolution of $\sim$18\arcsec\ (\citealt{price2001}). The MSX beam size is approximately 50 times smaller at 12 and 25 $\mu$m than that of the IRAS mission ($\sim$45\arcsec $\times$ 200\arcsec\ at 12 $\mu$m) and therefore avoids many of the confusion problems encountered by IRAS. In total 38 of the 45 southern  BRCs are located within the MSX survey regions -- the seven clouds outside the MSX region are indicated by an asterisk to the right of their catalogue name given in Table~\ref{tbl:brc_catalogue}.

These four bands together can help to provide information on the large scale interaction  between the OB stars and the surrounding molecular material.
The most useful MSX bands for this study are bands A and E centred at 8~$\mu$m and 23~$\mu$m respectively. The MSX band A covers a wavelength range of 6.8--10.8 $\mu$m, this includes emission from the unidentified infrared (UIR) bands at 7.7 and 8.6 $\mu$m, which has been widely attributed to polycyclic aromatic hydrocarbons (PAHs; \citealt{leger1984}). These molecules are excited by ionising photons from OB stars that manage to penetrate into the surface layers of molecular clouds surrounding HII regions, it is this layer of material that is referred to as a PDR (e.g., \citealt{urquhart2003}). As the 8~$\mu$m emission is dominated by these PAHs in photo-dominated regions this particular MSX band is extremely useful for identifying the edges of HII regions (\citealt{rathborne2002}).

The emission in the MSX band E  (spanning 18 to 25~\mum) is due to warm dust at $\sim$100~K, and is therefore very useful for tracing thermal emission and identifying deeply embedded objects such as ultra-compact (UC) HII regions and young stellar objects (YSOs).

Spitzer has a number of instruments that are able to image 
from 3.5--160~\mum; the IRAC instrument has four bands centred at 3.5, 4.5, 5.8 and 8~\mum\ and the MIPS instrument has three bands centred at 24, 70 and 160~\mum. The IRAC 8~\mum\ and MIPS 24~\mum\ bands are therefore comparable to MSX, but offer far superior resolution and sensitivity. Searching through the Spitzer archive we found data on a significant number of our sample, however, a full analysis of the available data are beyond the scope of this paper and we postpone a full analysis of this data to a subsequent paper (Urquhart et al. in prep.).\footnote{Spitzer observations of BRCs were not made in any systematic or uniform way, with many clouds observed as part of different General Observer programmes. Thus data are not available for the complete sample. In addition, some observational data sets are still within their proprietary period and currently unavailable to us.} We therefore limit our discussion of Spitzer data to sources where it lends some insight into the physical structure of a few clouds for which no MSX data are available (e.g., SFO 50 and SFO 52).

\subsection{Summary of recent radio observations} \label{sect:radio_summary}

In an earlier paper we reported a set of radio continuum observations towards the whole southern sample of BRCs made with the Australia Telescope Compact Array (\citealt{thompson2004b}). These observations identified radio continuum emission, which traces the distribution of ionised gas, towards the bright rims of 26 clouds consistent with the presence of an IBL. These observations also
resulted in the identification of four compact and ultra compact (UC) HII regions. Higher resolution follow-up observations, and molecular line mapping observations of a subset of these radio detections, have confirmed our initial findings and have led to the identification of another candidate UCHII
region within SFO 58 (\citealt{urquhart2004,urquhart2006a,urquhart2007d}). In a set of more recent high-resolution radio observations (Urquhart et al. in prep.) we have identified radio sources that are positionally coincident with the embedded thermal sources located behind the rims 
of another three clouds, and are thus good candidates to be UCHII regions. The results of these radio observations are summarised in the final column of Table~4.

\section{Results and Analysis}
\label{sect:results}

In this section we present the results and analyses of the CO and mid-infrared archival image data. To assist with the presentation of our results in the following subsections, we present a DSS image (left panels) and plots of the observed CO spectra (right panel) for each cloud in Fig.~9. We have overlaid contours of the 8~\mum\ emission on the DSS images to illustrate the correlation between the PDR and optically bright rims. The position of the IRAS point source and embedded 21~\mum\ MSX source (if detected; see Table~\ref{tbl:msx_ps}) associated with each cloud are indicated by a cross and triangle respectively.

In addition to the images and plots presented in Fig.~9  we present a summary of the physical parameters in Table~4 derived from the CO data, luminosities and other associated star formation tracers found in the literature. In Columns 2 and 8--11 we present the far-infrared luminosities taken from \citet{sugitani1994} (upper limits are indicated by parenthesis), associated MSX point source, non-Gaussian CO profiles, associated masers and radio continuum emission, respectively. In Columns 3--7 we present values derived from the CO data presented in this paper and estimates of the ionising fluxes. (The three clouds towards which no CO emission was detected have been omitted from this table.)

\renewcommand{\thefootnote}{\alph{footnote}}
\setcounter{footnote}{0}
\begin{table*}[!t] %
\begin{minipage}{\textwidth} 
\begin{center} 
\caption{\label{tbl:CO_derived_parameters} Summary of the physical parameters of the BRC sample including IRAS luminosities,  CO derived temperatures and densities, and other star formation tracers. }

\begin{tabular}{@{}l....c.clll@{}} 

\hline\hline
&\multicolumn{1}{c}{Lum.}& \multicolumn{1}{c}{$T_{12}$} & \multicolumn{1}{c}{$\tau_{18}$}& \multicolumn{1}{c}{$T_{13/18}$} &
\multicolumn{1}{c}{log $N$(H$_2)$} & \multicolumn{1}{c}{UV Flux} & \multicolumn{4}{c}{Tracers}\\ 

\cline{8-11} SFO Id. 	&\multicolumn{1}{c}{(L$_\odot$)}& \multicolumn{1}{c}{(K)} & 
& \multicolumn{1}{c}{(K)} & \multicolumn{1}{c}{(cm$^{-2}$)} &
\multicolumn{1}{c}{(10$^8$ cm$^{-2}$ s$^{-1}$)}&MSX PS&CO&Masers&Radio\\ \hline
\multicolumn{11}{c}{Spontaneous Sample\footnotemark \setcounter{footnote}{1}} \\
\hline
45&(30)&13.6&\multicolumn{1}{c}{$\cdots$}&\multicolumn{1}{c}{$\cdots$}&21.5&3.93&
\multicolumn{1}{c}{N}\\
46&20&14.1&\multicolumn{1}{c}{$\cdots$}&\multicolumn{1}{c}{$\cdots$}&21.0&1.46&
\multicolumn{1}{c}{$\cdots$}\\
47&(8400)&24.8&\multicolumn{1}{c}{$\cdots$}&\multicolumn{1}{c}{$\cdots$}&21.4&2.61
&\multicolumn{1}{c}{Y}& & & \\
48&(0.83)&8.1&\multicolumn{1}{c}{$\cdots$}&\multicolumn{1}{c}{$\cdots$}&20.8&1.84
&\multicolumn{1}{c}{$\cdots$}\\
49&(4300)&26.0&\multicolumn{1}{c}{$\cdots$}&\multicolumn{1}{c}{$\cdots$}&21.8&3.32
&\multicolumn{1}{c}{Y}& & & \\ 
50&(2.2)&6.2&\multicolumn{1}{c}{$\cdots$}&\multicolumn{1}{c}{$\cdots$}& $<$20.7&4.34&\multicolumn{1}{c}{$\cdots$}& & &  \\ 
51&16&13.3&0.25&10.3&22.0&6.99 &Y& Red wing&\\
52&22&17.9&0.24&12.1&22.0&1.39 &\multicolumn{1}{c}{$\cdots$}& Wings&&\\
53&(3.4)&17.5&0.20&11.6&21.9&1.47&\multicolumn{1}{c}{$\cdots$}\\ 
56&23&13.8&\multicolumn{1}{c}{$\cdots$}&\multicolumn{1}{c}{$\cdots$}&$<$20.6&1.13&\multicolumn{1}{c}{N}& & & \\
59&13000&14.9&\multicolumn{1}{c}{$\cdots$}&\multicolumn{1}{c}{$\cdots$}&21.4&36.79
&\multicolumn{1}{c}{N}& &  &HII region\\
60&(280)&9.8&\multicolumn{1}{c}{$\cdots$}&\multicolumn{1}{c}{$\cdots$}&$<$21.2&2529.35&\multicolumn{1}{c}{Y}\\
61&750&6.6&\multicolumn{1}{c}{$\cdots$}&\multicolumn{1}{c}{$\cdots$}&$<$20.9&0.16&\multicolumn{1}{c}{N}& & &\\
62&18000&23.9&\multicolumn{1}{c}{$\cdots$}&\multicolumn{1}{c}{$\cdots$}&21.7&16.07
&\multicolumn{1}{c}{N} & & & HII region\\

80&(220)&14.3&\multicolumn{1}{c}{$\cdots$}&\multicolumn{1}{c}{$\cdots$}&21.4&273.77
&\multicolumn{1}{c}{Y}\\
81&(38)&14.9&\multicolumn{1}{c}{$\cdots$}&\multicolumn{1}{c}{$\cdots$}&21.1&32.25&
\multicolumn{1}{c}{N}\\
83&(110)&4.9&\multicolumn{1}{c}{$\cdots$}&\multicolumn{1}{c}{$\cdots$}&$<$20.7&3.13&\multicolumn{1}{c}{N}& & &  HII region\\ 
86&880&24.8&0.20&11.1&22.2&7.12&N\\
&&8.4&0.45&7.0&22.0&\multicolumn{1}{c}{$\cdots$}&N\\

\hline 
\multicolumn{11}{c}{ Triggered Candidate Sample\footnotemark \setcounter{footnote}{1}} \\
\hline

57&(210)&32.1&0.07&18.9&22.2&43.48  &\multicolumn{1}{c}{N}& Red shoulder& &  PDR\\ 
58&140&34.6&0.07&20.2&22.1&16.15&\multicolumn{1}{c}{N}&Blue shoulder& &  PDR/UCHII\\ 
66&(190)&19.8&0.20&9.5&22.0&6.35&\multicolumn{1}{c}{N}&Wings?&H$_2$O/Methanol& PDR\\
67&(470)&29.7&0.20&12.4&22.3&6.75&\multicolumn{1}{c}{N}& Blue Shoulder&&PDR\\

70&490&18.2&\multicolumn{1}{c}{$\cdots$}&\multicolumn{1}{c}{$\cdots$}&21.3&6.7&\multicolumn{1}{c}{N}& & &PDR\\ 
73&(210)&16.7&0.08&9.8&21.8&18.81 &N\\
& &13.1&0.08&9.8&21.6&\multicolumn{1}{c}{$\cdots$}&N\\
84&450&20.5&\multicolumn{1}{c}{$\cdots$}&\multicolumn{1}{c}{$\cdots$}&21.5&2.24&\multicolumn{1}{c}{N}& & & PDR\\ 
87&(1400)&37.0&0.17&14.7&22.1&125.95&\multicolumn{1}{c}{N}\\
88&(2200)&30.0&\multicolumn{1}{c}{$\cdots$}&\multicolumn{1}{c}{$\cdots$}&21.8&99.04&\multicolumn{1}{c}{N}\\
89&(39)&11.3&\multicolumn{1}{c}{$\cdots$}&\multicolumn{1}{c}{$\cdots$}&20.9&14.16&
\multicolumn{1}{c}{N}& & & PDR\\ \hline
54&(1100)&22.5&0.24&12.1&22.5&1.05 & Y&Wings&Methanol&PDR/UCHII?\\ 
55&350&35.0&0.18&13.2&22.2&2.29
&Y& Red wing&H$_2$O&\\ 

64&14000&24.2&0.08&11.3&22.0&19.66&Y& Wings&H$_2$O/Methanol&PDR/UCHII\\
65&1100&18.8&0.17&7.9&22.0&4.88&Y& & & PDR/UCHII?\\
68&(3400)&25.4&0.19&12.9&22.4&18.39&Y&Wings&H$_2$O/Methanol&PDR\\
69&(1300)&20.0&\multicolumn{1}{c}{$\cdots$}&\multicolumn{1}{c}{$\cdots$}&21.3&21.16
&\multicolumn{1}{c}{Y}& &  &PDR\\ 
71&1500&20.2&0.19&11.7&22.3&\multicolumn{1}{c}{$\cdots$}&\multicolumn{1}{c}{Y}&Blue wing&H$_2$O&\\
72&(1800)&31.3&0.09&15.5&22.0&81.39&Y&Red wing&&\\

74&5500&26.2&0.01&29.8&22.0&5.42&Y& &&UCHII\\

75&34000&41.9&0.20&19.7&22.8&178.60&Y& &Methanol&PDR\\ 
76&5600&35.3&0.04&22.2&22.1&107.15&Y& &H$_2$O&PDR\\
79&(4400)&49.0&0.18&18.7&22.7&157.36&Y& & H$_2$O&UCHII\\ 
82&(1500)&31.6&0.17&14.2&22.4&1.08&Y& & & PDR\\
85&18000&36.0&0.05&11.4&22.7&1.08&Y&Red wing&Methanol&PDR/UCHII\\ \hline

\end{tabular}
\footnotetext[1]{This sample includes all BRCs that should no evidence that they are being photoionised and therefore any star formation occurring within them is unlikely to have been triggered.}
\footnotetext[2]{This sample includes all sources found to be associated with a PDR or IBL and are therefore thought to be good candidates for triggered star formation.}

\end{center}
\end{minipage}
\end{table*}

\renewcommand{\thefootnote}{\arabic{footnote}}
\setcounter{footnote}{4}

\subsection{CO results}
\label{sect:co_results}

We detect CO emission towards all but three BRCs (SFO 63, 77 and 78), with all three lines being detected towards just over half (23) of the 45 clouds observed. Multiple components are detected towards 16 BRCs, however, it is relatively easy to identify the component of interest associated with each BRC,
with the exceptions of 73 and 86; the spectra towards these display strong emission at two distinct velocities, with both having an equal probability of being associated with the BRC. To identify the correct component these regions will need to be observed using a high density tracer such as CS, or mapped in CO. However, since we are unable to identify a unique component we will include both possibilities.

Towards five clouds (SFO 50, 56, 60, 61 and 83) only the $^{12}$CO transition is detected. This isotopomer is the most abundant of the three observed. However, the non-detection of the rarer isotopomers would suggest that the molecular gas associated with these clouds is rather diffuse and of low column density. Assuming these, and the three non-detections (8 sources in total) identify a class of low density clouds we find this makes up a significant proportion of the sample ($\sim$18~\%). Furthermore, five of the eight candidate low-density clouds are located within the MSX region (SFO 56, 60, 61, 63 and 83) and are not associated with any embedded mid-infrared point sources (see Sect.~\ref{sect:msx_results} for details), which is consistent with the low-densities inferred from the CO emission. Interestingly, radio emission has been detected towards the rims of the three clouds that are CO non-detections, which could indicate that these clouds, due to a combination of low cloud densities and a strong radiation
field are being flash ionised. This possibility will be investigated in more detail in Sect.~\ref{sect:cloud_zapping}.

Comparing the peak velocities for the remaining 37 sources towards which we detect CO emission in two, or more transitions, reveals no significant velocity variations between transitions. However, examination  of the $^{12}$CO line profiles with respect to the other two transitions reveals 13 sources which display a significant deviation from the expected Gaussian shape. The CO profiles seen towards three clouds show evidence for the presence of a secondary component which appears as a shoulder on the observed profile. The presence of these secondary components could indicate the presence of shocked gas in the surface layers, as we might expect to find in clouds undergoing RDI. Alternatively, it is also possible that these secondary components are due to the superposition of another cloud along the line of sight at a similar velocity. 

In addition to the shoulder components high velocity wings are seen towards ten clouds. The presence of these wings may indicate the presence of a protostellar molecular outflow. Unfortunately, without mapping these sources we are unable to confirm the presence of an outflow which requires the spatial distribution of the red and blue wings to be traced. However, searching the literature we find four clouds are associated with Herbig-Haro (HH) objects (SFO 51, 52, 54 and 64; \citealt{sugitani1994}), a strong indication that an outflow is present. Furthermore,  we find an additional four of these sources are also associated with water masers (\citealt{valdettaro2007}; see Sect.~5.1 and Table~4 for more details). H$_2$O masers are widely thought to be associated with molecular outflows and/or accretion  (\citealt{elitzur1989,garay1999} also see review by \citealt{fish2007}). The detection of a wing in itself can only be taken as an indication that an outflow may be present, however, the fact that eight of the ten sources that exhibit wings are also associated with either a H$_2$O maser or HH object support our identification of molecular outflows.

\subsection{MSX analysis}
\label{sect:msx_results}

We obtained MSX band A images of all 38 BRCs (see Table~\ref{tbl:brc_catalogue} for details) for which data are available from the MSX archive (\texttt{http://irsa.ipac.caltech.edu/data/MSX/}). As previously mentioned we used these images to identify sources that are subject to photoionisation through the strong PAH emission produced in PDRs that dominates this band. By comparing the distribution of 8~$\mu$m emission with the morphology of the bright rim seen in the optical images we find a good correlation between the two towards 24 ($\sim$70\%). Comparing these clouds with those found to be associated with radio continuum emission we find 16 clouds in common -- in fact these 16 include every cloud towards which an IBL has been confidently identified. Only eight clouds are not associated with radio emission coincident with their bright rims. However, we find that our radio observations towards the majority of these were not sensitive enough to have detected the emission from the ionised gas (estimated from the predicted ionising photon flux incident on the surface of the clouds; see \citealt{thompson2004b}). 

In order to identify clouds that contain embedded sources such as compact HII regions or YSOs we searched the MSX point source catalogue (\citealt{egan2003}) for all potential sources within a 2\arcmin\ radius of the IRAS position. We discarded any MSX point sources not detected in the 21~$\mu$m band and examined the 21~$\mu$m emission maps to differentiate genuine point sources from diffuse thermal emission more likely to be associated with the PDR. In total we find MSX point sources towards 19 BRCs (50\% of the sample). The MSX point sources associated with each cloud and their mid-infrared fluxes are presented in Table~\ref{tbl:msx_ps}.  

\renewcommand{\thefootnote}{\alph{footnote}}
\setcounter{footnote}{0}
\begin{table*} %
\begin{minipage}{\textwidth} 
\begin{center} 
\caption{The properties of MSX point sources associated with the BRC sample.} \label{tbl:msx_ps}

\begin{tabular}{@{}lccc.....c@{}} \hline \hline & & & &   &\multicolumn{4}{c}{Mid-infrared fluxes} &
\\ \cline{6-9} SFO Id. & MSX name & RA & Dec. &  \multicolumn{1}{c}{Offset} & \multicolumn{1}{c}{8
$\mu$m} & \multicolumn{1}{c}{12 $\mu$m}& \multicolumn{1}{c}{14 $\mu$m}& \multicolumn{1}{c}{21
$\mu$m} & Quality flag\footnotemark \setcounter{footnote}{0}\\

&			&	(J2000)&(J2000)	   &  \multicolumn{1}{c}{(\arcsec)} & 
\multicolumn{1}{c}{(Jy)}&  \multicolumn{1}{c}{(Jy)}&  \multicolumn{1}{c}{(Jy)}& 
\multicolumn{1}{c}{(Jy)}& \\

\hline 47&G234.7617$-$00.2766& 07:31:49.0&  $-$19:27:34.2& 1.3	&2.5  & 3.5   &3.4	&11.9   &4444 
\\ 49&G234.6358+00.8281& 07:35:38.9&  $-$18:48:50.4& 27.0  &3.4  & 4.6   &3.3	&27.2   &4444\\
51&G253.2934$-$01.6114& 08:09:33.1&  $-$36:04:59.5& 3.8	&0.5  & 0.9   &1.2	&2.3    &4121\\
54&G259.9395$-$00.0419& 08:35:31.0&  $-$40:38:25.8& 8.5	&0.5  & 1.1   &3.4	&30.0   &4444\\
55&G260.7658+00.6604& 08:41:06.7&  $-$40:52:16.7& 72.4  &1.5  & 2.3   &1.9	&3.4    &4443\\
60&G268.0574$-$00.9515& 09:00:01.2&  $-$47:31:46.2& 9.4	&2.8  & 5.8   &2.3	&7.2    &4444\\
64&G290.3745+01.6615& 11:12:18.2&  $-$58:46:19.9& 2.0	&0.8  & 2.8   &10.0  &51.2   &4444\\
65&G294.2815$-$01.9124& 11:32:48.7&  $-$63:27:21.6 & 56.9  &0.2  & \multicolumn{1}{c}{$-$0.7}  &1.1	&4.2
&3023\\ 68&G294.5117$-$01.6205& 11:35:32.4&  $-$63:14:41.3& 10.4  &0.8  & 1.3   &2.5	&13.9   &4244\\
69&G295.1622$-$01.5825& 11:41:11.8&  $-$63:23:23.3& 11.1  &-0.0 & \multicolumn{1}{c}{$-$0.6}  &0.9	&2.8
&0022\\ 
71&G304.8872+00.6356& 13:08:12.2&  $-$62:10:21.0& 6.2	&0.1  & \multicolumn{1}{c}{$-$0.6} 
&1.1	&5.2    &1024\\ 
72&G306.1160+00.1386& 13:19:07.9&  $-$62:33:42.8& 1.9	&3.1  & 4.6  
&5.5	&10.1   &4444\\ 
74&G313.2850$-$00.3350& 14:19:42.0&  $-$61:25:12.0  & 5.0	&0.5  & 1.7  
&2.1	&8.5    &4344\\ 
75&G327.5761$-$00.8497& 15:55:50.6&  $-$54:38:47.4& 10.7  &4.6  & 5.9  
&2.3	&7.1    &4444\\ 
76&G332.9565+01.8035& 16:10:38.6&  $-$49:05:52.1& 1.0	&6.2  & 10.0 
&9.3	&22.4   &4444\\ 
79&G336.4917$-$01.4741& 16:40:00.0    & $-$48:51:41.4 &3.7    &58.2 & 147.3
&264.2 &1796.3 &4444\\
80&G336.6568$-$01.4099& 16:40:22.6&  $-$48:41:43.4& 69.1  &0.1  &
\multicolumn{1}{c}{$-$0.6}  &1.1	&4.7    &1023\\ 82&G343.0500+02.6094& 16:46:53.5&  $-$41:13:59.5& 39.6
&2.1  & 2.3   &1.5	&4.1    &4433\\ 85&G343.3516$-$00.0786& 16:59:07.0&  $-$42:42:07.2& 5.8	&74.4 &
75.4  &46.3  &119.0  &4444\\ 

\hline

\end{tabular} 
\footnotetext[1]{Each of the four numbers (starting on the left) are associated with the  the 8, 12, 14 and 21~\mum\ flux values and indicate the quality of each flux value as follows: 0 = not detected, 1 = upper-limit, 2 = fair, 3 = good, 4 = excellent.}

\end{center} 
\end{minipage} 
\end{table*}

\renewcommand{\thefootnote}{\arabic{footnote}}
\setcounter{footnote}{4}

\begin{figure} \setlength{\unitlength}{1cm}
\includegraphics[width=0.95\linewidth]{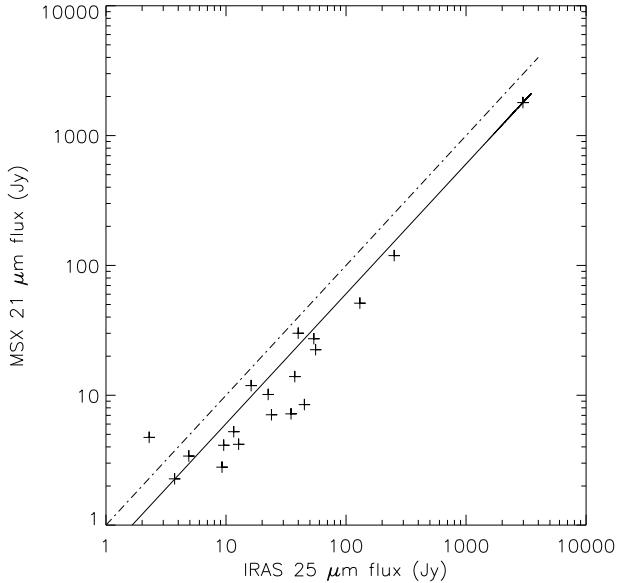}

\caption{\label{fig:msx_iras_flux} Plot comparing the 21~$\mu$m flux of MSX point sources identified towards BRCs with the 25~\mum\ flux of the IRAS point source associated with the same cloud. The flux measurements are shown as crosses and the linear least squared fit to the data is shown as a
continuous line. The dashed line indicates where the crosses would be if the MSX and IRAS fluxes where equal.}

\end{figure}

All of these MSX sources are located behind the bright rim towards the
interior of the cloud. The location of the 21~$\mu$m emission and its point-like appearance are consistent with the presence of an embedded object, which in turn is consistent with the hypothesis that star formation is taking place within them. Comparing the 21~\mum\ MSX fluxes of the 19 point sources to those measured by IRAS at 25~\mum\ we find them to be, in general, $\sim$50\% lower (see Fig.~\ref{fig:msx_iras_flux}), however, taking into account  
differences in wavelength and beam size between MSX and IRAS, the fluxes are in reasonable agreement.  If this trend were to continue towards longer wavelengths it would mean that the luminosities estimated from IRAS fluxes may have only been overestimated by a factor of a few. This compares well with the results reported by \citet{morgan2008} from SCUBA observations of a sample of northern BRCs. They found the luminosities obtained with the superior resolution of HiRes IRAS and submillimetre data were typically 2--3 times lower that those calculated by \citet{sugitani1991} from IRAS fluxes alone. 

\begin{figure} \setlength{\unitlength}{1cm}

\includegraphics[width=0.95\linewidth]{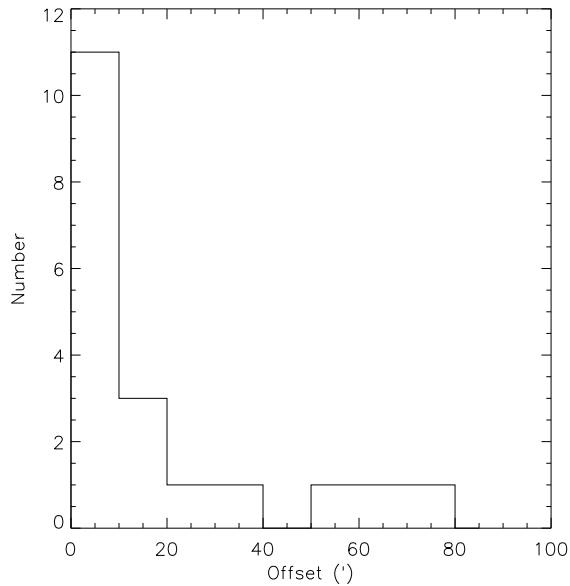}

\caption{\label{fig:msx_iras_offset} Plot comparing the positional offset between the IRAS and MSX point sources. The bin size used is 10\arcsec.}

\end{figure}

In Fig.~\ref{fig:msx_iras_offset} we present a histogram plot of the offset between the MSX and IRAS point source positions. This plot reveals that the positions of MSX and IRAS sources are fairly well correlated with each other. The majority are within 20\arcsec\ and given the size of the IRAS beam are again in good agreement. Since our observations were centred on the position of the IRAS point source associated with each cloud it means the derived parameters for the majority of the clouds will be a good indication of the embedded core properties. The offset between the positions for three clouds is larger than 40\arcsec, this  may indicate the presence of a second, colder, embedded source in addition to the warmer embedded source inferred from the presence of the MSX source. This raises the interesting possibility of star formation multiplicity within these clouds, possibly an indication that a number of star formation sites are present. Moreover, the presence of a mid-infrared bright core (i.e., detected in MSX) and a core only detected at far-infrared wavelengths is suggestive of different evolutionary stages within a cloud -- this would be expected if the star formation has been triggered by a shock wave traveling through the cloud i.e.,  sequential star formation. 

Comparing the clouds associated with PDRs and those with embedded thermal sources we find twelve in common. The presence of a PDR clearly indicates a significant interaction between the HII regions and the clouds. Furthermore, the presence of an embedded thermal source is a strong indication that star formation is taking place within these clouds. It is therefore possible the star formation within these 12 BRCs has been triggered, making them excellent candidates for further study. The absence of an embedded mid-infrared point source does not necessarily imply that star formation is not taking place within a cloud, as the star formation may still be in its infancy and has not yet progressed to a stage where it can be detected in the mid-infrared. For example, the SCUBA observations reported by \citet{morgan2008} detected submillimetre cores towards SFO 87 and SFO 89 that, as yet, do not have a mid-infrared counterpart. Alternatively, these  may be examples of clouds that have only recently been exposed to the ionisation front and have not yet had time for star formation to have been triggered. The twelve BRCs associated with PDRs but without an MSX point source may therefore be in an earlier evolutionary stage than the clouds with MSX point sources; this possibility will be discussed in more detail in Sect.~\ref{sect:evolutionary_sequence}.

Fourteen BRCs show no evidence of possessing a PDR which calls into question their initial selection as sites of triggered star formation. It is also worth noting that no radio emission has been detected towards the rims of any of these clouds. Five of these are associated with embedded sources, however, since there is no evidence of photoionisation these are more likely to be examples of spontaneous star formation rather than potential sites of triggered star formation via RDI. The observed bright-rims seen towards these clouds could be the result of illumination by nearby low-mass stars. Alternatively these clouds may be located too far from the ionising star for the radiation field to significantly affect their evolution. 

\subsection{The Gum Nebula}

\begin{figure*}
\begin{center}
\includegraphics[width=0.45\linewidth]{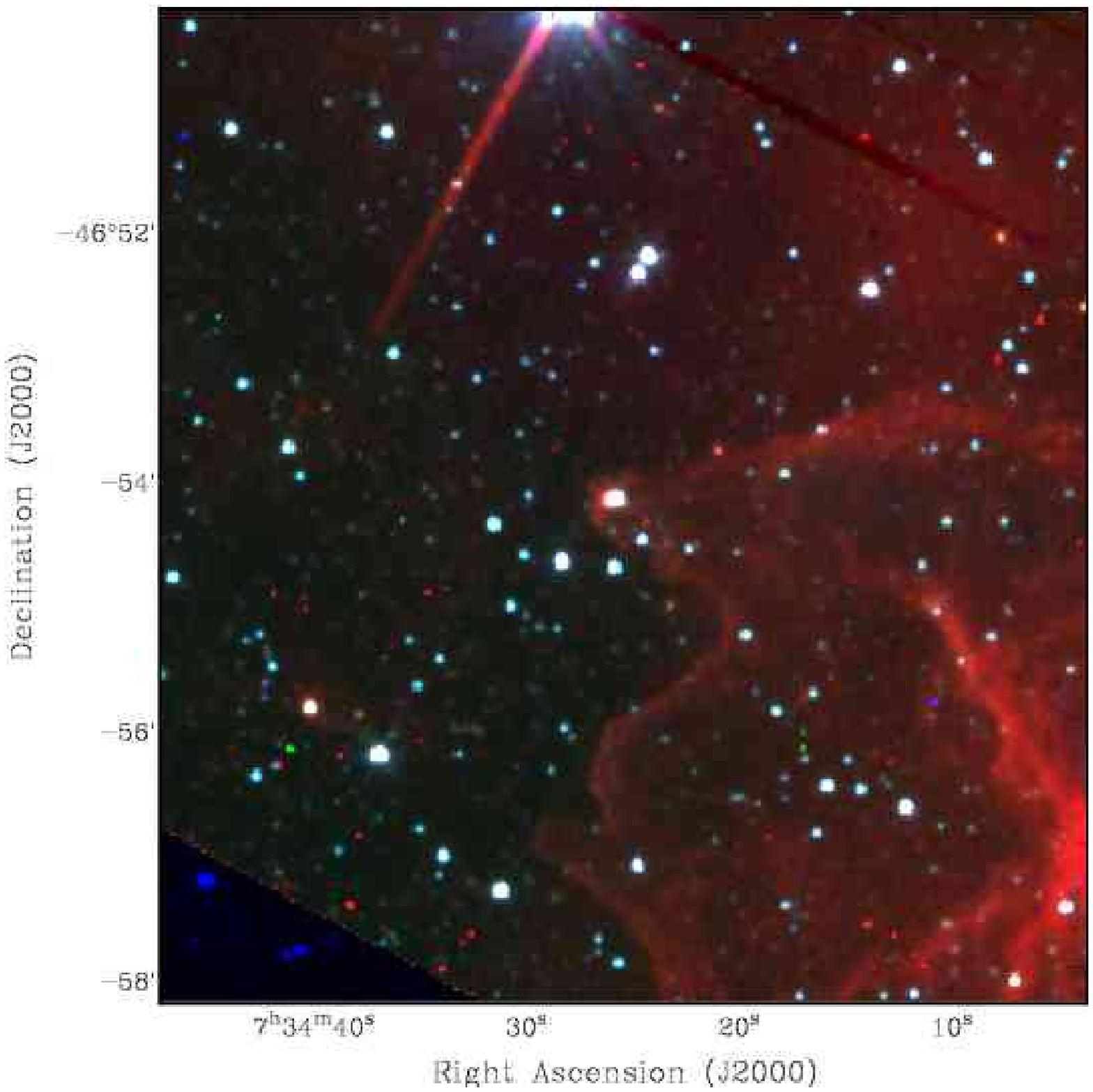}
\includegraphics[width=0.45\linewidth]{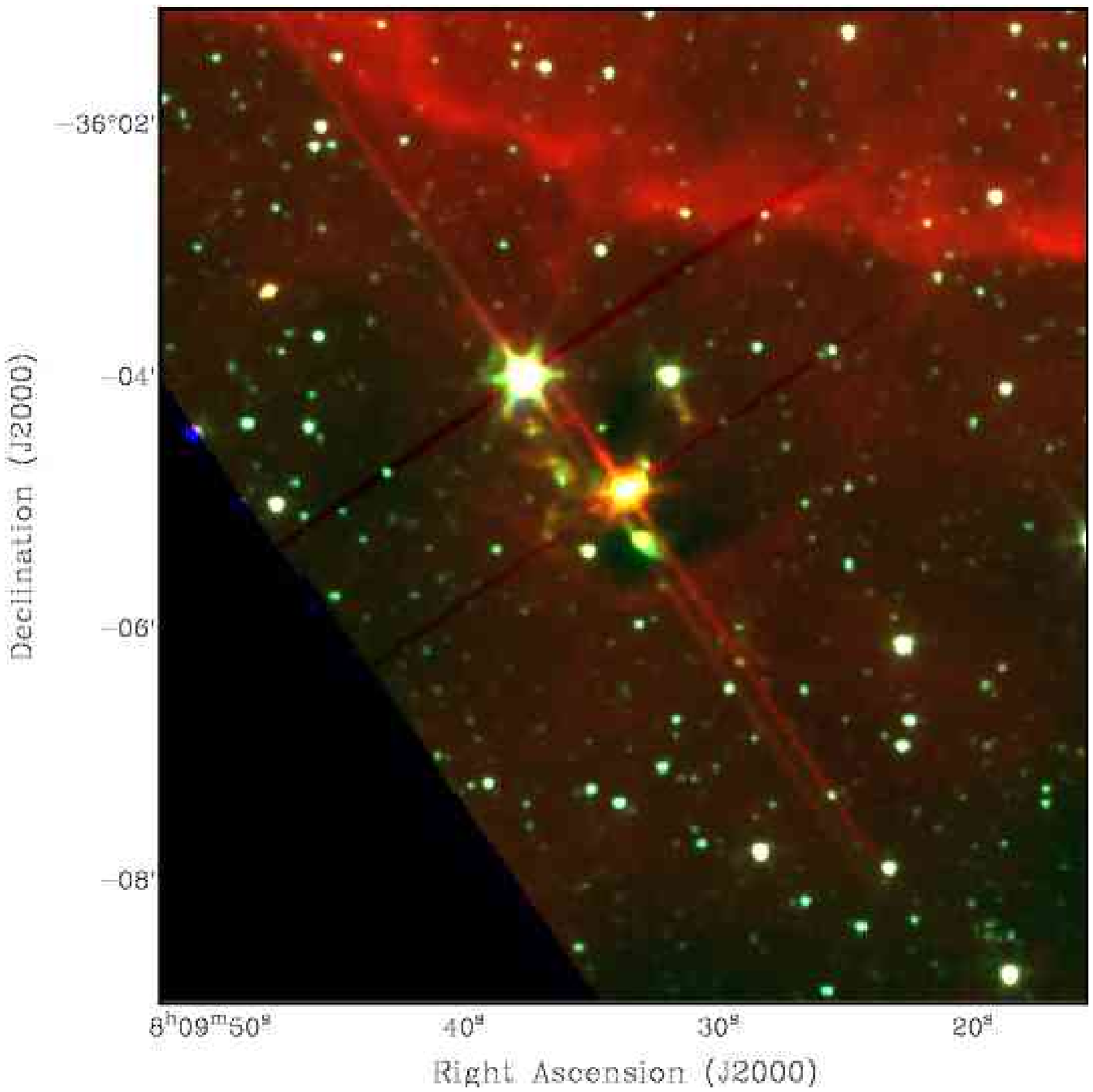}

\caption{\label{fig:gum_irac} Three colour images of the BRCs SFO 48 and  SFO 51 left and right panels respectively. These images have been produced by combining the 3.6, 4.5 and 8~\mum\ IRAC bands which are shown in blue, green, and red, respectively. Note the diffuse green emission in SFO 51.}

\end{center}
\end{figure*}

In the previous subsection we used MSX data to separate a sample of 38 BRCs into two groups; those likely to have been triggered and those in which triggering is unlikely. However, the lack of any MSX data for the BRCs associated with the Gum Nebula has not permitted us to determine which of these two groups they fall into. The Gum Nebula is located at a heliocentric distance of $\sim$450~pc and covers a region of 30\degr$\times$30\degr\ (\citealt{sugitani1994} and references therein). It is excited by two stars, $\zeta$ Pup and  $\gamma^{2}$ Vel, which have spectral types O4 If and WC 8+O8 III respectively (\citealt{yamaguchi1999}). There are six BRCs associated with Gum nebula (SFO 46, 48, 50-53), only one of which is located within the MSX survey region (SFO~51). 

Although this HII region is relatively nearby, its size, and the location of its associated BRCs, result in very large projected distances between the ionising stars and the cloud rims. The projected distances range from 32~pc to 91~pc. At these distances the upper limit to the photon flux incident at the surface of these clouds is relatively low ranging from $\sim$1.4--$7\times 10^8$~\photons\ with an average of only a few times $10^8$~~\photons. These are among the smallest photon fluxes incident on any BRCs in the southern sample. It is therefore not surprising to find that no radio emission is detected towards any of these clouds ($<$ a few mJy at 20 and 13~cm; \citealt{thompson2004b}).

Three of these six BRCs have been observed using  IRAC  on  Spitzer (SFO~48, SFO~51 and SFO~52). Spitzer images have recently been presented by \citet{velusamy2008} of SFO~52 and its associated outflow and so will not be reproduced here. In Fig.~\ref{fig:gum_irac} we present three colour composite images of the other two clouds (SFO~48 and SFO~51)  produced by combining the 3.5, 4.5 and 8~\mum\ IRAC bands. As with the 8~\mum\ MSX band, the 8~\mum\ IRAC band is dominated by emission from PAHs and is therefore a good tracer of PDRs. In these images the 8~\mum\ band is coloured red to make the identification of any associated PDRs relatively straightforward. However, looking at these images (the two presented in Fig.~\ref{fig:gum_irac} and the image of SFO~52 (\citealt{velusamy2008})) we fail to find strong 8~\mum\ emission towards any of these three clouds that might indicate the presence of a PDR. This is consistent with the low photon flux estimated on their surfaces and the non-detection of any radio emission. 

There is strong evidence that star formation is occurring within at least two of the Gum Nebula BRCs. Both SFO 51 and SFO 52 are associated with HH objects. Furthermore, the composite IRAC image of SFO 51 presented in Fig.~\ref{fig:gum_irac} displays a diffuse region of 4.5~\mum\ excess emission, which is commonly referred to as a \emph{green fuzzy} or an \emph{extended green object} (\citealt{cyganowski2008}). This excess emission is associated with shocked gas, as both the H$_2$($v$=0--0, S(9,10,11)) lines and CO ($v$=1--0) bandheads lie within the IRAC 4.5 \mum\ passband \citep{cyganowski2008}, and is thought to indicate the presence of a molecular outflow (e.g., \citealt{beuther2007}). In addition, the $^{12}$CO spectra detected towards these two sources both show evidence of high velocity wings which again indicates the presence of molecular outflows.

In spite of the evidence that star formation is taking place within a couple of the BRCs associated with the Gum Nebula, there is no evidence to suggest that any of these clouds are being photoionised. From the observational data available we therefore conclude that the current star formation is unlikely to have been induced by RDI. We conclude that these BRCs are more likely to be examples of spontaneous star formation and have grouped these with the group of clouds that were not found to be associated with a PDR in Sect.~\ref{sect:msx_results}.

\subsection{CO non-detections: a case of cloud zapping?}
\label{sect:cloud_zapping}

There are three BRCs that have not so far been discussed, these are the BRCs towards which no CO emission is detected (i.e., SFO~63, SFO~77 and SFO~78). It is interesting to note that radio emission has been detected towards all three of these clouds (\citealt{thompson2004b}). The correlation between the morphology of the radio emission and the optical rims of these clouds supports the presence of an IBL, clearly demonstrating a strong interaction is taking place between these clouds and the HII region. The non-detection of molecular emission and the presence of strong radio emission might imply that these clouds are being flash photoionised.  An alternative explanation may be that all of the molecular material towards these BRCs has been photodissociated and what we are seeing in the optical images is the interface between the HII and HI regions (cf. \citealt{deharveng2006}). Observations of either the HI or mid-infrared emission from these clouds to trace their atomic gas or PDRs are required to unambiguously determine their nature. These data are not yet currently available. 

\begin{figure*}[!t]
\setlength{\unitlength}{1cm}
\begin{picture}(15,12)

\put(-0.5,0.2){\includegraphics[width=0.72\linewidth]{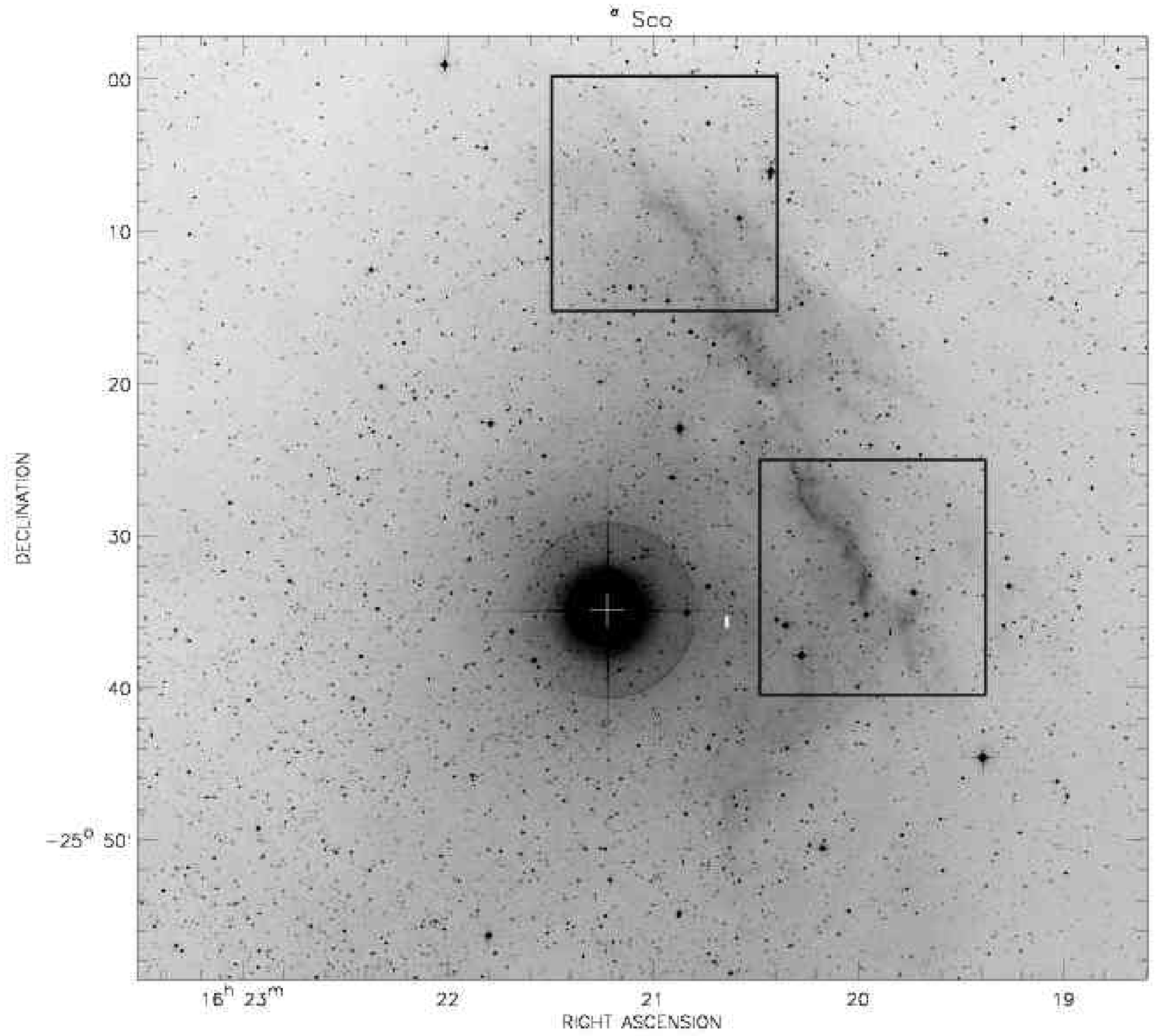}}
\put(12,6.5){\includegraphics[width=0.31\linewidth]{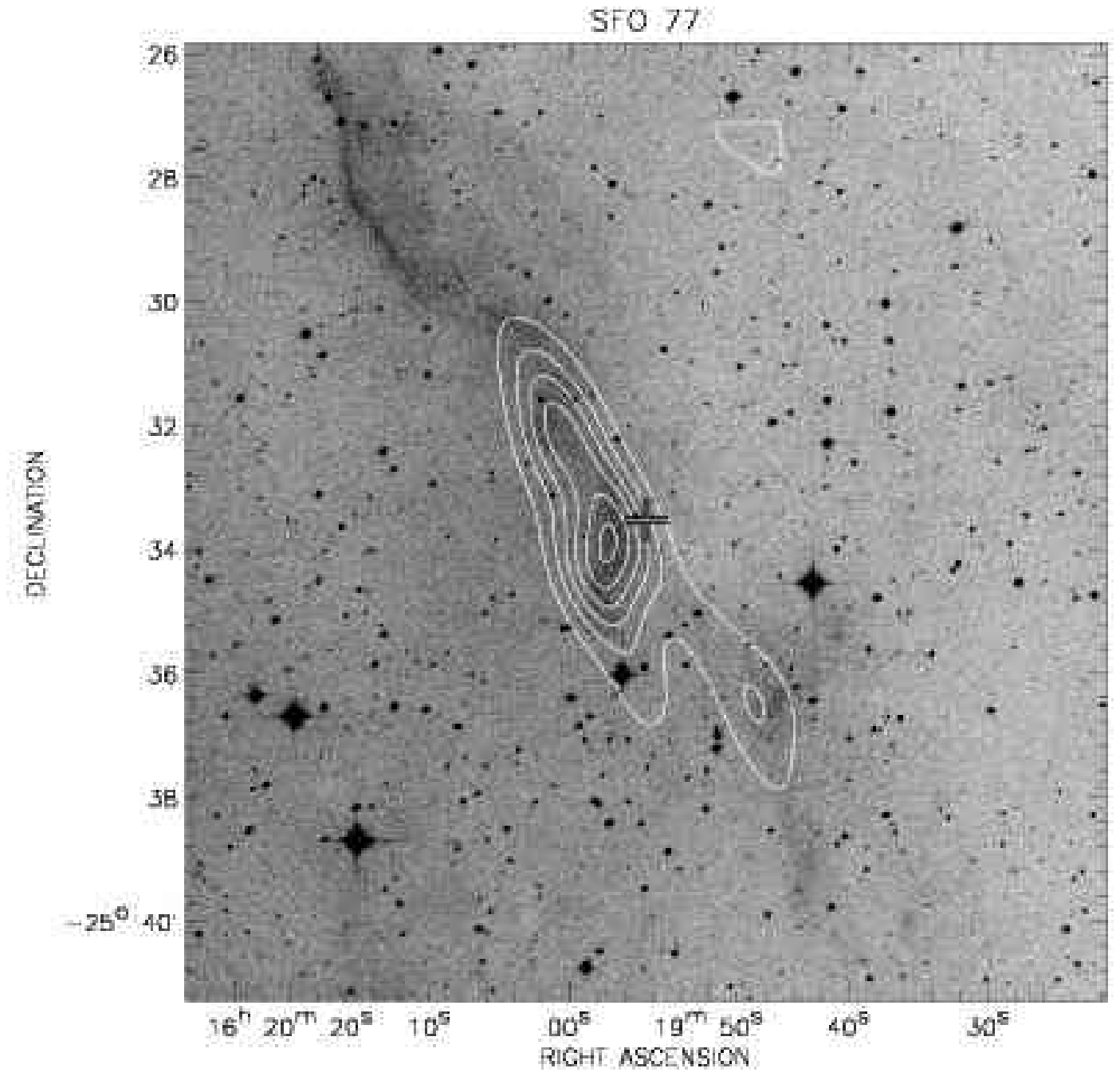}}
\put(12,1.25){\includegraphics[width=0.31\linewidth]{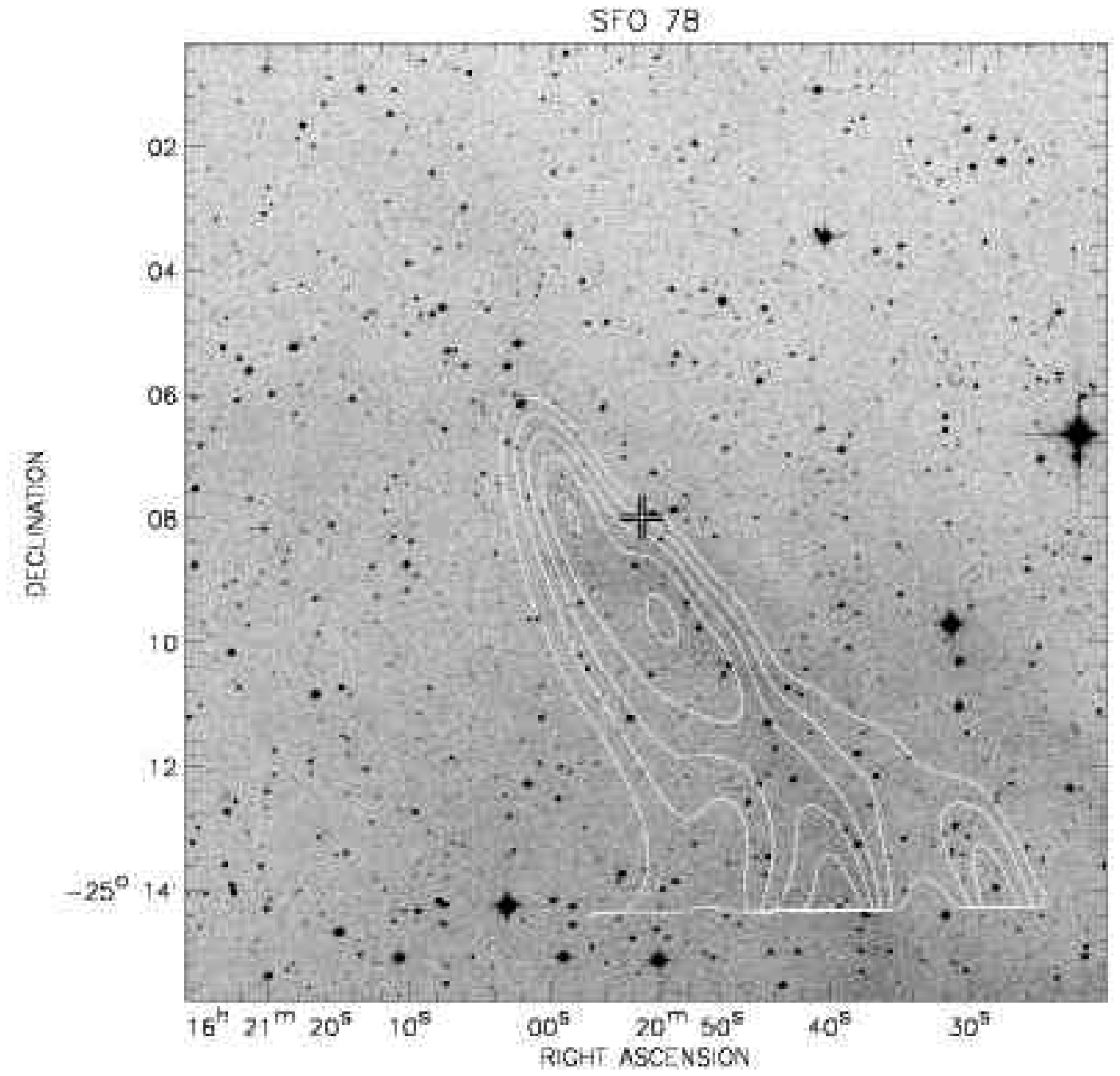}}

\end{picture}\vspace{-40pt}
\caption{\label{fig:sigma_sco}  Left panel: Large scale DSS image of $\sigma$~Sco. In this image we indicate the locations of the two BRCs associated with this HII region by outlining them in black. Right panels: Enlarged images of the region around SFO 78 (upper panel) and SFO 77 (lower panel) overlaid with contours of the radio continuum emission detected towards them (see \citealt{thompson2004b} for details). Crosses in these two images indicate the position of the IRAS sources associated with the clouds.}

\end{figure*}

The radio emission is particularly strong towards the rims of SFO~77 and SFO~78 which are both associated with $\sigma$~Sco HII region. At a  distance of only 150~pc $\sigma$~Sco  is therefore the closest of the HII regions in our sample. In the left panel of Fig.~\ref{fig:sigma_sco} we show a large scale image of the HII region and its surrounding regions. The HII region is excited by two B1 stars (HD 147165 and CCDM J16212--2536AB) the positions of which are indicated in the image by crosses. The two BRCs associated with $\sigma$~Sco are both located within a long ridge of bright emission which runs in a north-eastern direction, beginning approximately due north of the central stars and ending to the west. We present close up images of these two BRCs in the two panels presented on the right panels of Fig.~\ref{fig:sigma_sco}. The position of the IRAS point sources associated with the BRCs are indicated by crosses in each of these images.

The bright ridge of visible-wavelength emission lies at a projected distance of $\sim$~0.7-1~pc from the ionising stars. Given the ridge's close proximity to the ionising stars we would expect it to be exposed to an intense UV flux. It is therefore unsurprising to have detected 3 and 6 cm radio continuum emission associated with this ridge ($\sim$1--10$\times$10$^{8}$~photon~cm$^{-2}$ s$^{-1}$); \citealt{thompson2004b}). The non-detection of any CO emission towards either of these clouds  suggests that the ridge is very low column density, supported by the non-detection of any submillimetre emission towards either cloud (\citealt{morgan2008}).

The intense radiation field and the inferred low column densities of the molecular material associated with all three BRCs leads us to conclude that it is highly likely these clouds are being rapidly destroyed by an ionisation flash. These clouds are therefore not examples of BRCs in which star formation is currently, or is ever, likely to take place.

\section{Discussion}
\label{sect:discussion}

In the previous section we presented the results of our CO observations as well as MSX and Spitzer analysis. We are now able to separate all 45 BRCs into three distinct groups: 1) clouds not associated with a PDR,  any star formation is therefore unlikely to have been the result of RDI (18 BRCs including the 6 clouds associated with the Gum Nebula, or $\sim$40\% of the southern BRC catalogue); 2) clouds associated with a PDR/IBL and therefore possible sites of triggered star formation (24 clouds $\sim$53\%); 3) diffuse clouds that are being ionised by strong radiation fields and may be examples of flash ionisation (3 clouds $\sim$7\%). 

In this section we will concentrate on the first of these two groups which we will hereafter refer to as spontaneous and triggered samples respectively, where the name indicates the most likely mode of star formation. Our discussion will focus on two important questions: 1) are there fundamental differences between the properties of the triggered and spontaneous samples? 2) Do the clouds in the triggered sample with, and without, an embedded MSX point represent distinct evolutionary stages?

In Table~\ref{tbl:CO_derived_parameters} we summarise the result of the CO observations  presented in Sect.~\ref{sect:co_results}. In addition, we indicate any non-Gaussian profile seen in the CO data, whether any H$_2$O or methanol masers have been detected, and finally whether radio emission has been detected. Where radio emission has been detected we distinguish between compact radio emission likely to be associated with an embedded object (e.g., UCHII region) and more extended emission that is coincident and morphologically correlated with the rim of a cloud (i.e., an IBL). To help facilitate the discussion to follow we separate the BRCs into their respective groups (i.e., spontaneous and triggered candidates). Furthermore, we subdivide the triggered candidates into two groups depending on whether or not they are associated with an embedded MSX point source.  

\subsection{Comparisons between sites of triggered and spontaneous star formation}
\label{sect:comparisons}

\subsubsection{Physical parameters}

\begin{figure*}
\begin{center}
\includegraphics[width=0.4\linewidth]{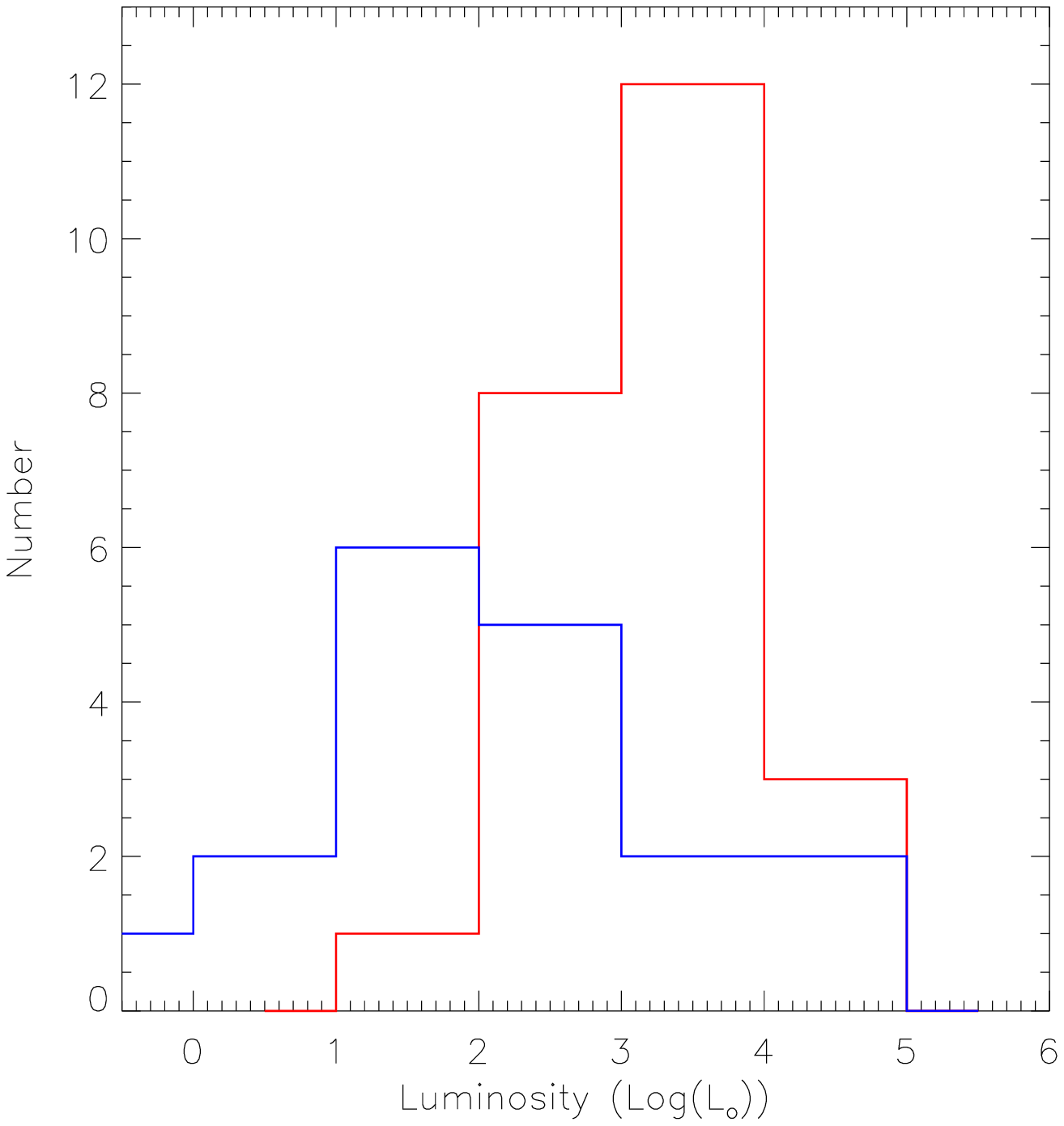}
\includegraphics[width=0.4\linewidth]{1149fig6a.eps}\\
\includegraphics[width=0.4\linewidth]{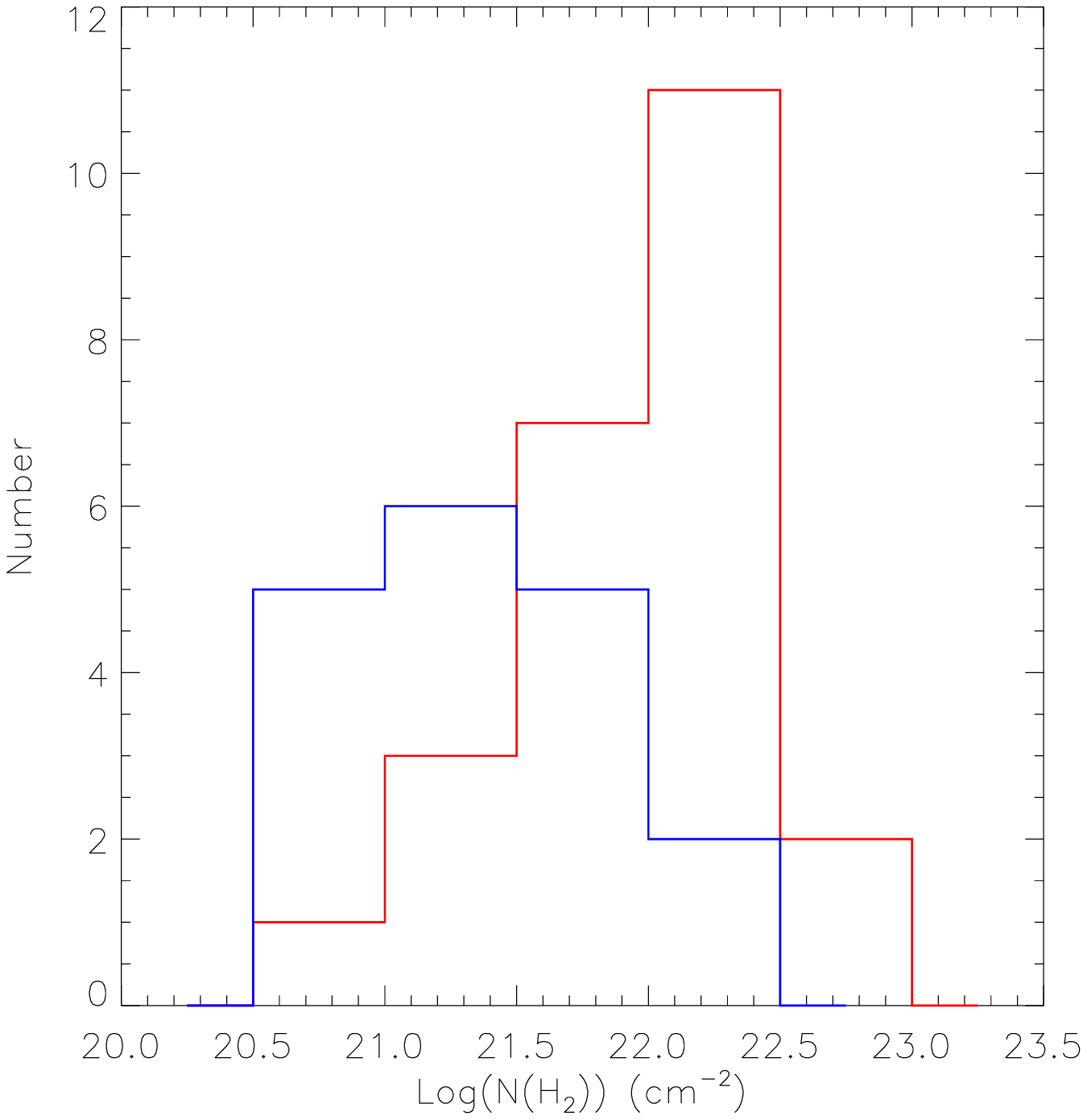}
\includegraphics[width=0.4\linewidth]{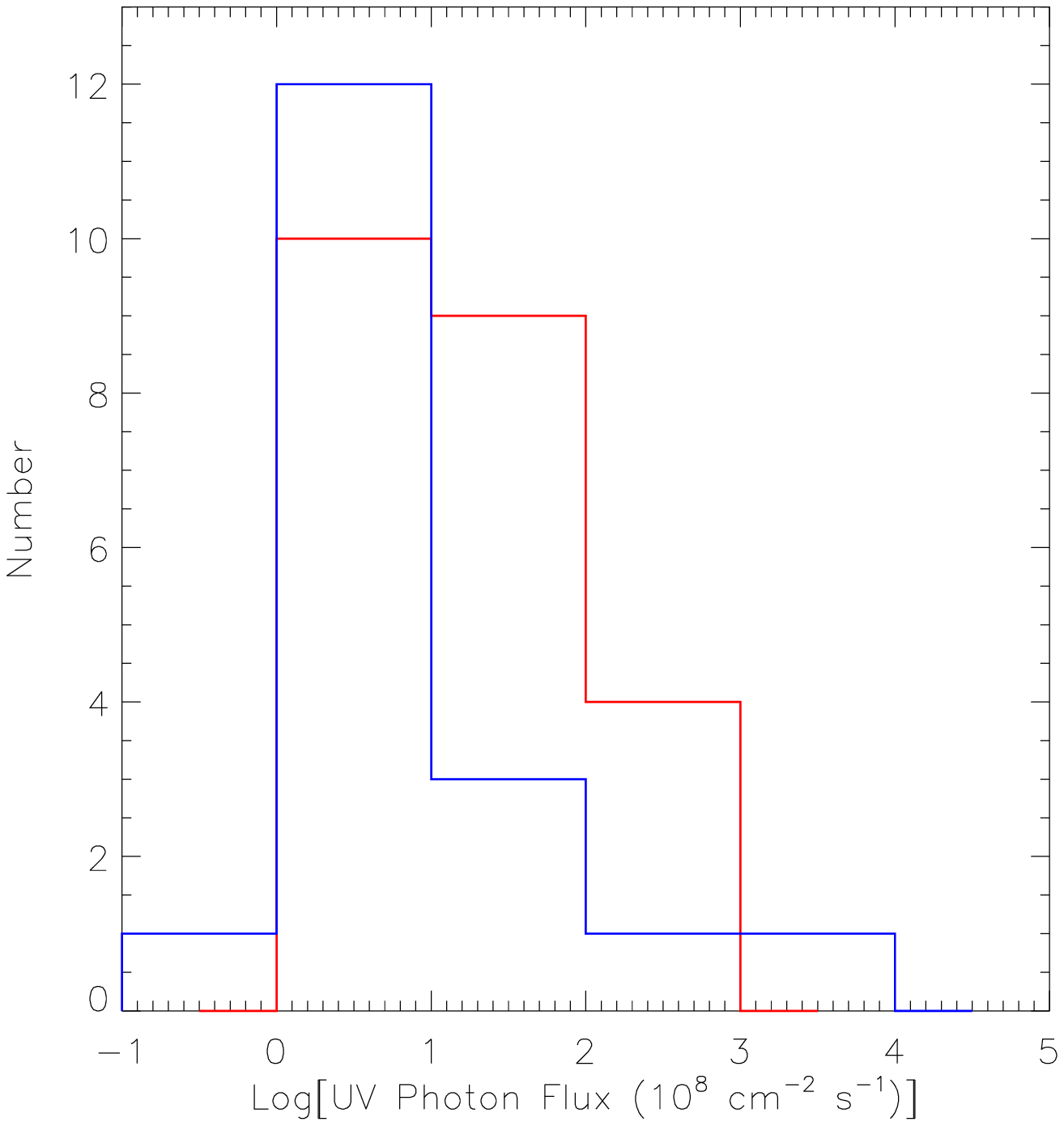}

\caption{\label{fig:comparison_plots} Histogram plots comparing the luminosities, surface temperatures, column densities and estimate of the ionising photon flux between the clouds identified as triggered and spontaneous star formation sites (outlined in red and blue respectively). 
}
\end{center}
\end{figure*}

Comparing the parameters of the triggered and spontaneous candidate clouds tabulated in Table~\ref{tbl:CO_derived_parameters} we can immediately see some striking differences. In Fig.~\ref{fig:comparison_plots} we present histogram plots of the luminosity, surface temperature, column densities and ionising flux distributions of the triggered and spontaneous clouds which clearly illustrate these differences between the two groups. The three most obvious are the luminosities of the embedded IRAS sources, the surface temperatures (i.e., excitation temperature derived from the $^{12}$CO peak intensity) and column densities of the clouds, all of which appear to be significantly enhanced in the sample of triggered clouds.

When comparing the intrinsic luminosities of the bright-rimmed clouds Ê
in our sample we must also consider the fact that the clouds are Ê
illuminated by a strong UV flux which may be responsible for a Ê
proportion of a clouds luminosity due to external heating of the Ê
clouds. We calculated upper limits to the UV heating for each cloud Ê
using the luminosities of the illuminating stars and simple geometry Ê
to constrain the flux that could be absorbed by each cloud and re- 
radiated. In the majority of cases the incident UV illumination can be Ê
responsible for no more than a few percent of the total luminosity, Ê
with the exception of four clouds (SFO 48, 60, 80 and 81). All of Ê
these clouds are from the spontaneous sample.

The average luminosity of sources embedded within our sample of triggered clouds is $\sim$1000~\lsun, which is an order of magnitude larger than the average luminosity calculated for the source embedded within the spontaneous clouds ($\sim$100~\lsun). It is also worth noting that the spontaneous sample contains a number of UCHII regions (i.e., SFO 59 and SFO 62; \citealt{thompson2004b}) that have been incorrectly identified as BRCs, and stellar clusters (SFO 47 and SFO 49; \citealt{messineo2007}). All four of these clouds have luminosities of several thousand \lsun, and thus skew the average luminosity to a higher value. If we exclude these anomalous sources we obtain an average luminosity for the spontaneous sample of a couple of tens of \lsun, which is more representative of the sample as a whole. These luminosities are similar to protostars associated with more isolated clouds and Bok globules, consistent with our conclusion that any star formation seen towards these clouds is more likely to be spontaneous than triggered.

Having eliminated the clouds in which star formation is considered unlikely to have been triggered, we find we have also excluded nearly all of the embedded sources with luminosities under a couple of hundred \lsun. As mentioned in the introduction, the luminosities of IRAS point sources associated with BRCs are found to be systematically more luminous than IRAS point sources associated with regions of isolated clouds. Although this analysis relies on a number of luminosities calculated using IRAS 60 and 100 \mum\ upper limits, it does give additional credence to the suggestion that this mode of star formation may lead to the formation of more massive stars, and to an overall enhancement of the  star formation efficiency.  

 In Col. 3 of Table~4 we present the $^{12}$CO excitation temperature (T$_{12}$), determined from the emission assumed to be optically thick; we consider this to probe the surface conditions of the cloud. The surface temperature distributions of the spontaneous and triggered samples are very different, with the former peaking between 10--15~K -- only slightly larger than would be expected from heating of the interstellar radiation field. The lower temperatures found toward these sources is consistent with the non-detection of any radio or mid-infrared emission. The surface temperatures found towards the triggered sample are significantly higher  with an average value of $\sim$30~K, which is twice that found towards the spontaneous sample. This difference in surface temperatures between the two samples is consistent with our hypothesis that surfaces of the triggered sample of clouds are being externally heated through the photoionisation of their surface layers.

\begin{figure}
\begin{center}
\includegraphics[width=0.95\linewidth]{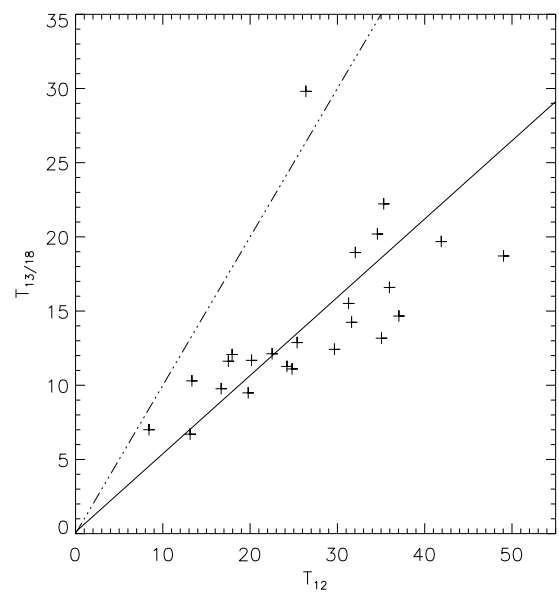}

\caption{\label{fig:temperature_correlation} Plot comparing the temperature of the gas as derived
from the peak $^{12}$CO emission to the excitation temperature estimated from less optically thick
$^{13}$CO and C$^{18}$O transitions. The least-squared best fit to these data is shown as a solid line and the line of equality is indicated by the dashed-dotted line. }

\end{center}
\end{figure}

For the 22 clouds towards which all three isotopomers are detected we have used the moderately optically thick $^{13}$CO and optically thin C$^{18}$O transitions as a probe of their internal structures. In Cols. 4--5 of Table~\ref{tbl:CO_derived_parameters} we present the estimated values for the optical depth and internal temperature of these clouds. It would be interesting to compare these derived parameters for the two groups, but unfortunately only four of these are members of the spontaneous sample -- too small a number to obtain any reliable statistics. Since the clouds we suggest to be triggered are exposed to a more intense radiation field, as inferred from the presence of a PDR and ionised gas around their rims,  we might expect to find higher temperatures in the surface layers, traced by the optically thick transition of $^{12}$CO, than their interiors, traced by the moderately optically thick $^{13}$CO and optically thin C$^{18}$O. In Fig.~\ref{fig:temperature_correlation} we present a plot of the difference between the external ($T_{12}$) and internal ($T_{13/18}$) temperatures for all 22 clouds; the straight line shows the least squared best fit to these data. This figure nicely illustrates the difference of excitation conditions between the $^{12}$CO and the optically thin isotopes. We find that the external temperatures are, in general, a factor of two larger than the internal temperatures. This significant temperature gradient between the clouds' surface layers and their interiors is consistent with the assumption 
that the clouds are exposed to a significant amount of external heating.

Column densities have been calculated using the optical depth and internal excitation temperature derived from the optically thin transition of C$^{18}$O and the moderately optically thick transition of $^{13}$CO (i.e., $\tau_{18}$ and $T_{13/18}$) for clouds where all three transitions are detected. For sources where only the $^{12}$CO and $^{13}$CO are detected we have used the excitation temperature derived from the optically thick $^{12}$CO to determine the optical depth and hence their column densities, and for clouds where only the $^{12}$CO is detected we estimate an upper limit for the column density using a value of 3$\sigma$, where $\sigma$ is standard deviation of the noise measured from the $^{13}$CO spectra. (For a more detailed discussion on how these quantities are derived see \citet{urquhart2006a} and references therein.) These column densities are presented in Col.~6 of Table~4 and the column density distributions of the triggered and spontaneous samples are shown in the lower left panel of Fig.~6. Looking at this plot it is clear that the column densities found towards the potentially triggered sample are systematically higher than found towards the spontaneous sample. The enhanced column densities found towards the triggered sample may have resulted from the compression of the cloud by the ionisation front.

In order to give an indication of the range of ionisation conditions present across the sample  we present estimated values of the UV flux incident on the surface of these clouds in Col.~7 of Table~4. These fluxes have been calculated using the fluxes values tabulated by \citet{panagia1973} for the ionising stars given in Table~1 and assuming that the projected distances between the cloud and its ionising star(s) are close to the actual distances. These estimated fluxes are strict upper limits since the actual distance can be much larger than the projected distances, and we have neglected any attenuation due to recombination and/or UV absorption, both of which could significantly reduce these values. For example, in their detailed study of their detailed study of the Trifid Nebular \citet{lefloch2002} found that the measured flux was almost an order of magnitude lower than predicted. In the lower right panel of Fig.~6 we present a plot of the ionising fluxes incident on the surfaces of the triggered and spontaneous samples. The distributions are clearly different with the triggered sample being associated with more intense radiation, however, the difference is not as striking as for the other three parameters discussed so far -- a reflection of the higher uncertainly associated with this parameter.

The four histograms presented in Fig.~6 reveal physical differences between the clouds within our triggered and spontaneous samples which support our separation criteria. In an effort to see whether any of these differences are statistically significant we conducted Kolmorgorov-Smirnov (KS) tests on all four parameters (i.e., luminosity, surface temperature, column density and ionising flux). We found we are able to reject the null hypothesis that the two distributions are drawn from the same population with 99.9\% or higher confidence level for the luminosity, surface temperature, and column density distributions, respectively. (For comparison a 3$\sigma$ difference corresponds to 99.87\%.) The rejection of the null hypothesis for the ionising flux distributions was found to be at the 94\% confidence level, which is below the 3$\sigma$ level and so cannot be dismissed. The results of the KS tests for luminosity, surface temperature and column density serve to confirm that these two samples really do represent statistically different cloud populations.

\subsubsection{Star formation tracers}

In Col.~8 of Table~\ref{tbl:CO_derived_parameters} we comment on any deviation of the observed CO profiles from the expected Gaussian shape (see Sect.~\ref{sect:co_results}). In total 13 clouds display non-Gaussian profiles, which as previously mentioned may indicate of the presence of a shock compression or molecular outflow, and although it is not possible to investigate this further with the present data, these secondary components and wings do imply these clouds are dynamically active. We note that in all but two cases these non-Gaussian profiles are associated with the triggered clouds, clearly demonstrating their dynamic nature. The low number of non-Gaussian profiles detected towards clouds in the spontaneous sample is a further indication that star formation may not have yet begun in the majority of these clouds. 

Methanol and H$_2$O masers are both excellent tracers of star formation. In general, Class II methanol masers are radiatively pumped and are thought to arise from inside an accretion disc, H$_2$O masers are collisionally pumped and are thought to be associated with molecular outflows (\citealt{fish2007}). Methanol masers are widely considered to be one of the earliest tracers of star formation and have the advantage of being almost exclusively associated with high-mass protostars (\citealt{walsh2001}). H$_2$O masers have been found in regions of high- and low-mass star formation and therefore act as a good general sign post of star formation. The \citet{valdettaro2007} search of the southern SFO catalogue detected H$_2$O maser emission towards 7 clouds; due to the inherent variability of H$_2$O masers this is a lower limit. In addition to H$_2$O masers we searched the methanol maser catalogue of \citet{pestalozzi2005} and found three clouds coincident with Class II maser emission.  Three more methanol masers have recently been detected in the vicinity of BRCs by the Methanol Multibeam (MMB) survey (Gary Fuller priv. comm.; see also \citealt{green2008}). This brings the total number of BRCs associated with methanol masers to six. Since neither the \citet{pestalozzi2005} catalogue\footnote{The \citet{pestalozzi2005} catalogue has been compiled from all the methanol masers reported in the literature and includes many targeted observations and small survey regions.} or the MMB survey ($|b| < 2$) provides complete coverage of our sample of BRCs the number of methanol masers so far detected is also a lower limit.

A total of seven H$_2$O and six methanol masers have been found in the vicinity of ten BRCs, a very strong indication that star formation is taking place within these clouds. Comparing the distribution of these masers we find them all associated with our sample of triggered clouds, no masers are reported towards the spontaneous sample.    

Another comparison we can make is to look at the distribution of radio emission which traces the ionised gas associated with a cloud's IBL. As with the other tracers we have considered, we find a striking difference between the two samples. Radio emission has been detected towards 16 of the 24 clouds identified as having a PDR, and therefore likely to be undergoing a strong interaction with their HII regions. The non-detection of radio emission towards the other eight clouds is presumably due to poor sensitivity of the our initial observations (\citealt{thompson2004b}). No radio emission is detected towards any of the spontaneous sample with the exception of three clouds that are identified as compact HII regions (SFO 59, SFO 62 and SFO 83).\footnote{The radio emission detected towards SFO~83 was initially identified as being associated with an IBL by \citet{thompson2004b}, however, re-examination of the radio and mid-IR data leads us to conclude that the emission is more likely to be associated with a central source rather than associated with the cloud's rim.}  

In addition to the radio emission associated with the bright rims we detect compact radio emission coincident with the embedded mid-infrared sources towards seven clouds. These are thought to be UCHII regions, another indication that star formation is taking place, and perhaps more significantly,  that a large fraction of OB stars are being formed -- this is also supported by the presence of the methanol masers. Again these compact radio sources are exclusively detected towards the triggered sample of BRCs.

Taken in isolation the results from our low-resolution radio continuum observations (\citealt{thompson2004b}) are hard to interpret, particularly the low number of detections ($\sim$50\% of the observed sample). However, looking at the available evidence we find strong support for the separation of the clouds into the potentially triggered and spontaneous clouds. The triggered sample are systematically more luminous and has have higher surface temperatures and column densities than the spontaneous sample.  Furthermore, we find strong evidence for star formation within the triggered sample including  methanol and H$_2$O masers, embedded mid-infrared point sources and UCHII regions. In addition we find circumstantial evidence for star formation in the form of CO wings which may indicate the presence outflows. In total we find evidence of star formation towards 16 BRCs of the triggered sample ($\sim$66\%) compared to only two of the spontaneous sample ($\sim$11\%).

\subsection{Evidence of an evolutionary sequence?}
\label{sect:evolutionary_sequence}
In Sect.~\ref{sect:comparisons} we investigated the differences between clouds that show strong evidence of being photoionised and clouds that appear to be relatively unaffected by their nearby HII regions. We have found convincing evidence that these two samples are indeed distinct, which supports our classification of the clouds. Now having excluded all of the relatively uninteresting BRCs, at least from a triggered star formation perspective, we are left with 24 BRCs that are excellent candidates to further investigate the RDI mode of star formation. 

In order to compare the global properties of the potentially triggered clouds with those of the spontaneous clouds, all of the BRCs associated with a PDR/IBL were grouped together. However, as we discussed earlier, these clouds naturally separate into two groups; those associated with, and those not associated with, an embedded MSX point source (see Table~\ref{tbl:CO_derived_parameters} for details). The non-detection of an MSX point source towards ten clouds may indicate that they are in an earlier stage of their evolution; either they have only recently been exposed to the HII region, or any ongoing star formation within their boundaries is still in the earliest stages, when the majority of the luminosity is emitted at (sub)millimetre wavelengths.

\begin{figure}
\begin{center}
\includegraphics[width=0.95\linewidth]{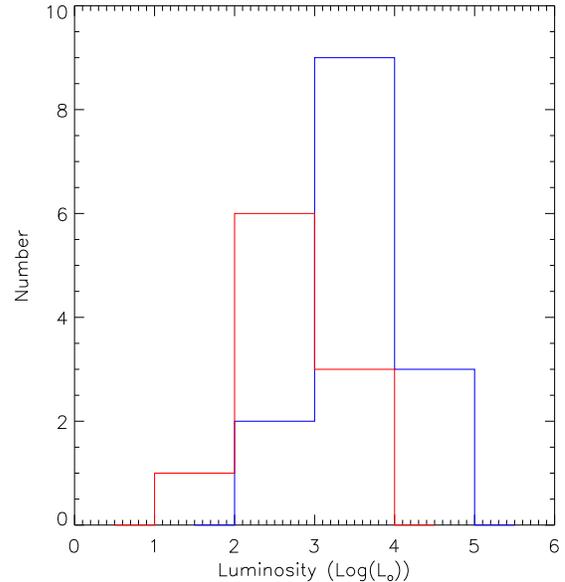}

\caption{\label{fig:histo_triggered} Histogram plot comparing the luminosities between the potentially triggered clouds with and without embedded mid-infrared point sources (outlined in blue and red respectively). 
}
\end{center}
\end{figure}

The detection of submillimetre cores embedded within two clouds (i.e., SFO 87 and SFO 89; \citealt{morgan2008}) not associated with an MSX point source supports this hypothesis.\footnote{\citet{morgan2008} used the SCUBA instrument on the JCMT to observe 44 BRCs. Only 6 of their sample are also included in our southern catalogue (SFO 45, SFO 77, SFO 78, SFO 87, SFO 88 and SFO 89) and consequently we are unable to assess the submillimetre contents of the whole sample.} In addition to the embedded submillimetre cores detected towards SFO~87 and SFO~89, a ridge of emission was detected towards the bright rim of SFO 88. The observed submillimetre emission is due to a combination of dust temperature and column density. The coincidence of the emission along the length of the bright rim may be due to heating by the radiation field, or could be due to density enhancement brought about by the compression of the cloud by the ionisation front. In either case the distribution would suggest that it has only recently been exposed to the HII region. We suggest that the triggered sample without a mid-infrared point source contains clouds at a variety of evolutionary stages, ranging from just after a cloud is exposed to the ionisation front, to just before the protostars become visible in the mid-infrared. It is therefore useful to compare the physical properties and tracers of these two groups in order to test this possible evolutionary sequence. We might expect to see differences between the two groups in luminosity, morphology, CO derived parameters, and maser associations. In Figure~\ref{fig:histo_triggered} we present a histogram comparing the luminosities of the two groups. This figure clearly demonstrates a trend towards higher luminosities for the clouds associated with MSX point sources. This is consistent with the expectation that luminosities will increase as the cloud evolves due to continual compression of the cloud by the ionisation front and accretion onto the developing protostars.

Sugitani et al. (1991, 1994) divided the catalogue into three morphological types depending on the curvature of their rims: (1) type A rims are moderately curved, (2) type B are tightly curved, and (3) type C are cometary in appearance. They suggested that these morphologies may represent an evolutionary sequence with clouds evolving from type A to C under the influence of the ionisation front. This idea has received some support from the numerical models of \citet{lefloch1994} and an observational study of six northern BRCs reported by \citet{devries2002}. Comparing the morphologies of the clouds associated with, and unassociated with an MSX point source, we find no evidence of any significant differences in morphology between them. However, the morphological evolution under the influence of RDI is strongly dependent on the initial properties of the cloud. The evolution of a cloud after the impact of the ionisation front can be highly non-linear with the final fate of the imploding cloud not easily predicted (\citealt{kessel2003}).  Any conclusions based upon morphological differences are therefore premature. 

There are no significant differences in the temperature or column density within the triggered sample, and we do not identify any significant trends between clouds with and without associated mid-IR sources. It is a different matter for the $^{12}$CO line profiles, however. In Sect.~\ref{sect:comparisons} we showed that nearly all of the sources that exhibit non-Gaussian profiles are associated with the potentially triggered sample. Now comparing how these are distributed with respect to the triggered clouds with and without embedded MSX point sources we find another clear difference. The three clouds identified as having shoulder components are associated with clouds without a MSX point source, and all but one of the sources that exhibit wings are associated with an MSX point source. The two types of non-Gaussian profiles observed arise from different physical processes: 1) shoulder components are the result of supersonic motions within the cloud, possibly the result of shock compression of the cloud;  2) wings are widely attributed to molecular outflows, high-velocity molecular gas thought to be linked to ongoing accretion, and therefore are an indication that star formation is taking place. 

These two processes may be understood in an evolutionary context with all shoulder components being associated with the earlier evolutionary stage when the ionisation front first begins to compress the cloud, presumably greatest in the early stages of RDI and, indeed, star formation. The presence of wings within so many of the clouds with an embedded mid-infrared point source, would certainly suggest these are more evolved than clouds without an embedded source. This conclusion is further supported by the fact that all but one of the detected masers and all but one of the potential UCHII regions are associated with the clouds with MSX point sources. 

The lower luminosities and lack of any evidence for current star formation within the vast majority of the clouds not associated with an MSX point source leads us to conclude that these clouds are more likely to be in an earlier stages of RDI than those that are associated with an MSX source.

\section{Summary and conclusions}
\label{sect:summary}

We present the results of a set of position-switched CO observations made towards all 45 bright-rimmed clouds from the Sugitani et al. (1994) southern catalogue. These observations were centred on the positions of the IRAS point sources  within each cloud. Three CO transitions were observed simultaneously ($^{12}$CO, $^{13}$CO and C$^{18}$O $J$=1--0) with CO emission being detected towards all but three clouds. We complement our CO observations with mid-infrared data obtained from the MSX and Spitzer archives.

Our main conclusions are as follows:

\begin{enumerate}

\item Analysis of the CO, mid-infrared and radio data has allowed us to separate the 45 BRCs which make up the SFO southern catalogue into three distinct groups; candidate spontaneous ($\sim$40\% of the sample), triggered ($\sim$53\%), and zapped  ($\sim$7\%) clouds, respectively: 

\begin{enumerate}
\item The 18 spontaneous clouds show no evidence for either photon-dominated regions (PDRs) or ionised boundary layers (IBLs) and are thus relatively unaffected by the photoionisation from the nearby OB star. These clouds are either located too far from their ionising stars for the radiation to significantly affect their evolution, or the illumination of their rims is due to radiation from nearby low-mass stars.
\item The 24 triggered clouds are found to be associated with a PDR which demonstrates a significant interaction between the molecular material and the HII regions.
\item The three clouds towards which no  $^{12}$CO emission was detected are all associated with strong ionisation fronts. These may be examples of clouds undergoing an ionisation flash.
\end{enumerate}

\item Comparing the physical parameters of the spontaneous and triggered samples we find striking differences in luminosity, surface temperature and column density with all three being significantly enhanced for the clouds considered to have been triggered. Furthermore, we find strong evidence for star formation within the triggered sample including  methanol and H$_2$O masers, embedded mid-infrared point sources and UCHII regions. In addition we find circumstantial evidence for star formation in the form of CO wings which may indicate the presence of an outflow. However, we find evidence of ongoing star formation within only two of the spontaneous sample.

\item Of the 24 BRCs identified as candidate triggered clouds we find that fourteen are also associated with embedded mid-infrared point sources. Towards these clouds we also find strong evidence that star formation is currently taking place. The non-detection of any tracers of star formation towards clouds not associated with a mid-infrared point source leads us to conclude that they represent an earlier evolutionary stage.  We find therefore that only half of the clouds undergoing RDI are currently forming stars, with the other half still in the very early stages of ionisation driven collapse.

\item The luminosities of the triggered clouds associated with an embedded mid-infrared point source are an order of magnitude larger than those unassociated with an embedded source, each having an average luminosity of a few thousand and a few hundred \lsun\ respectively. The large luminosities found towards the star forming clouds, the presence of methanol masers and detection of UCHII regions indicates that a significant number of OB stars are being formed. The formation of so many high-mass stars gives added support to the hypothesis that RDI leads to the formation of more massive stars and a higher star formation efficiency.

\item  We find that although the SFO catalogue contains a number of good examples of BRCs in which the star formation may have been triggered this only amounts to approximately 50\% of the sample. Moreover, from the current data it appears that only about half of this number is currently undergoing star formation (i.e., fourteen BRCs of a total of 45 in the sample $\sim$30\%). This is the main reason why previously reported results from observations have been extremely difficult to interpret.

\end{enumerate}

This work has refined the catalogue of \citet{sugitani1994} using the archival data sets of several legacy surveys. Potential triggered star-forming regions have been identified via a combination of millimetre spectroscopy, radio and mid-infrared imaging. By eliminating those sources that do not show true potential for an associated, photo-ionisation induced, triggering mechanism, we have produced 24 sources which are solid candidates for the production of RDI-induced triggered YSOs. These 24 candidates form an ideal basis for  future investigations into the RDI process of triggered star-formation. Also, given the wide-area nature of the archival data that we have used, we have shown that is eminently possible to construct a new larger catalogue of triggered star-formation sites. Through such an effort the true extent and effect of triggered star-formation in the Galaxy may be ascertained.

\begin{acknowledgements}

The authors would like to thank the Director and staff of the Paul Wild Observatory, Narrabri, New South Wales, Australia for their hospitality and assistance during our observations. We would also like to thank the referee for some useful comments and suggestions. JSU is supported by STFC and CSIRO  postdoctoral grants and LKM is supported, in part, by the NSERC. This research would not have been possible without the SIMBAD astronomical database service operated at CDS, Strasbourg, France and the NASA Astrophysics Data System Bibliographic Services. This research makes use of data
products from the MSX Survey, which is a joint project of the University of
Massachusetts and the Infrared Processing and Analysis Center/California Institute of Technology, funded by the National Aeronautics and Space Administration and the National Science Foundation.

\end{acknowledgements}

\bibliography{1149}

\bibliographystyle{aa}

\end{document}